\documentclass[nohyper,12pt,letterpaper]{JHEP3}
\usepackage[dvips]{epsfig}
\usepackage{amsfonts,amssymb}

\newcommand{\be}{\begin{equation}}
\newcommand{\ee}{\end{equation}}
\newcommand{\bea}{\begin{eqnarray}}
\newcommand{\eea}{\end{eqnarray}}
\newcommand{\ba}{\begin{array}}
\newcommand{\ea}{\end{array}}

\def\bbox{{\,\lower0.9pt\vbox{\hrule \hbox{\vrule height 0.2 cm
\hskip 0.2 cm \vrule height 0.2 cm}\hrule}\,}}
\newcommand{\dsl}{\pa \kern-0.5em /}

\newcommand{\EQ}{\begin{equation}}
\newcommand{\EN}{\end{equation}}

\def\bbox{{\,\lower0.9pt\vbox{\hrule \hbox{\vrule height 0.2 cm
\hskip 0.2 cm \vrule height 0.2 cm}\hrule}\,}}

\newcommand{\pa}{\partial}

\newcommand{\la}{\lambda}

\def\be{\begin{equation}}
\def\ee{\end{equation}}
\def\ba{\begin{eqnarray}}
\def\ea{\end{eqnarray}}
\def\la{~\mbox{\raisebox{-.6ex}{$\stackrel{<}{\sim}$}}~}
\def\ga{~\mbox{\raisebox{-.6ex}{$\stackrel{>}{\sim}$}}~}
\def\bq{\begin{quote}}
\def\eq{\end{quote}}

\makeatletter
\def\vereq#1#2{\lower3pt\vbox{\baselineskip1.5pt \lineskip1.5pt
\ialign{$\m@th#1\hfill##\hfil$\crcr#2\crcr\sim\crcr}}}
\makeatother  at 10truept

\newcommand{\beq}{\begin{equation}}
\newcommand{\eeq}{\end{equation}}
\newcommand{\beqa}{\begin{eqnarray}}
\newcommand{\eeqa}{\end{eqnarray}}

\def\la{~\mbox{\raisebox{-.6ex}{$\stackrel{<}{\sim}$}}~}
\def\ga{~\mbox{\raisebox{-.6ex}{$\stackrel{>}{\sim}$}}~}

\def\ltap{\ \raise.3ex\hbox{$<$\kern-.75em\lower1ex\hbox{$\sim$}}\ }
\def\gtap{\ \raise.3ex\hbox{$>$\kern-.75em\hbox{$\sim$}}\ }
\def\gl{\ \raise.5ex\hbox{$>$}\kern-.8em\lower.5ex\hbox{$<$}\ }
\def\roughly#1{\raise.3ex\hbox{$#1$\kern-.75em\lower1ex\hbox{$\sim$}}}

\title{Locally Localized Gravity: The Inside Story}

\author{Nemanja Kaloper\\
Department of Physics, University of California, Davis, CA 95616, USA \\
and \\
IHES, 35 Route de Chartres,
F-91440 Bures-sur-Yvette, France\\
E-mail: \email{{\tt kaloper@physics.ucdavis.edu}}}

\author{Lorenzo Sorbo\\
Department of Physics, University of California, Davis,
CA 95616, USA \\
E-mail: \email{{\tt sorbo@physics.ucdavis.edu}}}

\abstract{We derive the exact gravitational field of a
relativistic particle localized on an $AdS$ 3-brane, with
curvature radius $\ell$, in $AdS_5$ bulk with radius $L$. The
solution is a gravitational shock wave. We use it to explore the
dynamics of locally localized tensor gravitons over a wide range
of scales. At distances below $L$ the shock wave looks exactly
like the $5D$ $GR$ solution. Beyond $L$ the solution approximates
very closely the shock wave in $4D$ $AdS$ space all the way out to
distances $\ell^3/L^2$ along the brane. At distances between $L$
and $\ell$, the effective $4D$ graviton is a composite built of
the ultralight mode and heavier gravitons, whereas between $\ell$
and $\ell^3/L^2$ it is just the ultralight mode. Finally beyond
$\ell^3/L^2$ the shock reveals a glimpse of the fifth dimension,
since the ultralight mode wave function decays to zero at the rate
inherited from the full $5D$ geometry. We obtain the precise
bulk-side formula for the $4D$ Planck mass, defined as the
coupling of the ultralight mode, in terms of the $5D$ Planck mass
and the curvature radii. It includes higher-order corrections in
$L/\ell$, and reduces to the RS2 formula in the limit $\ell
\rightarrow \infty$. We discuss $AdS/CFT$ interpretation of these
results, and argue that the spatial variation of the effective
gravitational coupling read from the shock wave amplitude
corresponds to RG running driven by quantum effects in the dual
$CFT$. }


\preprint{ \tt{hep-th/0507191} \\
\\ }

\begin{document}

\section{Introduction}
\label{intro}

\subsection{Prelude}

The models of localized gravity in $AdS$ space \cite{rs2} offer a
new possibility for deriving $4D$ gravity at large distances from
a higher dimensional theory. Unlike in the conventional
Kaluza-Klein (KK) approach, the extra dimensions could be infinite
\cite{rs2,addk,ori}. However, when the boundary brane is Minkowski
\cite{rs2,addk,ori} or de Sitter \cite{ori,bent}, the bulk volume
is still finite. This yields a normalizable $4D$ zero mode
graviton, whose dynamics is described by $4D$ General Relativity
(GR). In addition, one also finds a continuum of massive KK modes.
It comes about because the boundary conditions far in the bulk are
arbitrary, and so the boundary value problem is one-sided. In a
certain sense these models might be thought of as
``semi-compactifications". Despite the KK continuum, gravity below
the cutoff appears $4D$ all the way down to infinity. This is
because the light KK modes couple to the brane probes
exponentially weakly, since they must climb the bulk gravitational
well to get to the brane, which in perturbation theory appears as
the familiar volcano potential suppression of \cite{rs2}. This
theory admits $AdS/CFT$ interpretation as a dual $CFT$ with a
cutoff coupled to 4D gravity \cite{cft,adscft}.

On the other hand, in the case of an $AdS_4$ 3-brane embedded in
$AdS_5$, or $AdS_4 \subset AdS_5$ for short, the bulk volume is
infinite \cite{bent,details}. The exact $4D$ zero mode graviton
decouples since the scale which controls its coupling, set by the
bulk volume, diverges. However, there arises a new light spin-2
mode with a mass $m \sim \ell^{-1} \arcsin(L/\ell)$, where $L$ and
$\ell$ are 5- and 4-dimensional $AdS$ radii, respectively
\cite{kara1,misa}. It is normalizable, and when $\ell \gg L$, it
is anomalously light, $m \sim L/\ell^2$, resembling the $4D$ zero
mode graviton. It plays a key role in the locally localized
gravity scenario advocated in \cite{kara1,kara0,kara2}. However,
it is not completely clear how to compute the $4D$ Planck mass
$M_4$ in locally localized gravity. Numerical integration
\cite{kara1} of the ultralight mode wave function suggests that
$M^2_4 \sim M^3_5 L$, qualitatively as in RS2 \cite{rs2}. The
arguments of \cite{bura} based on covariant entropy bound and
light-like holography should yield a finite answer, relating $M_4$
in some way to $M_5$ and the finite volume of the brane's
holographic domain. However \cite{bura} do not give an explicit
expression for this relation. Formulating $AdS_4 \subset AdS_5$ as
a dual effective $CFT$ coupled to $AdS_4$ gravity cut off at
$1/L$, \cite{porrati} computes the Planck scale to
be\footnote{This is the correct formula with normalizations as
found in \cite{rs2,addk}.} $M_4^2 = M^3_5 L$, precisely as in RS2
\cite{rs2}. Yet in light of \cite{bura} one might be circumspect
about using $AdS/CFT$ picture in the far infra-red. The dual
description based on the idea of defect $CFT$ \cite{defect} offers
a very nice picture, whereby a fat defect generates a small
anomalous dimension of $CFT$ operators that yield an ultralight
graviton, but the calculations involve strong coupling dynamics
and are not under complete control yet. Thus it is very
interesting to find a more precise rule for the calculation of
$M_4$. It is also interesting to find out whether gravity changes
form from $4D$ to $5D$, and where, if at all. In \cite{kara1}, it
has been suggested that at distances $\ga m^{-1} \sim \ell^2/L$,
the $AdS_4$ curvature screens out the deviations from the $4D$
gravity law. In \cite{bura} it was further argued that gravity may
remain $4D$ at all spatial scales for experiments that do not last
longer that a time $\Delta t \sim \ell$, but a more precise
statement is still lacking. These issues are puzzling, and
resolving them will shed more light on the mechanism of locally
localized gravity.

\subsection{Summary of the Main Results}

In this work we are able to explicitly calculate, on the bulk
side, the precise expressions for the $4D$ Planck scale $M_4$ and
the IR length scale at which gravity starts feeling the fifth
dimension. We derive the exact solution for the gravitational
field of a relativistic particle on the brane, a ``photon", which
generalizes the Aichelburg-Sexl shock wave of $4D$ GR
\cite{pirani,aichsexl} to $AdS_4 \subset AdS_5$ braneworlds. This
is the first example of an exact gravitational field of a particle
in locally localized gravity. It includes contributions from all
the tensor gravitons in the bulk spectrum, which contains three
sectors: the ultralight graviton with mass $m \sim L/\ell^2$, a
tower of ``intermediate" gravitons with masses $m \sim n/\ell$,
and the sector of ``heavy" gravitons with masses $m
> 1/L$ \cite{kara1,misa}. The couplings of heavier gravitons are
suppressed relative to the ultralight one by $\frac{L}{\ell} mL$,
similar to RS2 \cite{rs2,tunneling}, eventually saturating at
$\frac{1}{M_5^3\ell}$ for $m \gg L^{-1}$. We analyze the solution
in the regime $1/M_5 \ll L \ll \ell$ in order to separate the
scales that control the dynamics of the theory in a clear way. The
intricacies of the spectrum result in the emergence of {\it four}
interesting regimes of scales:

\begin{itemize}

\item[{{\bf I}\,:}]  ${\cal R} \ll L$: At
distances much shorter than the $AdS_5$ radius $L$, the solution
looks $5D$ as dictated by the short distance singularities of the
longitudinal Green's function. Because the momentum transfer
between the source of the gravitational field and a probe particle
is so high ($ > 1/L$) all the graviton modes contribute, and in
particular the very heavy ones dominate since they outnumber the
light modes and their couplings are less suppressed, and their
Yukawa-suppressions are negligible in this regime.

\item[{{\bf II}\,:}] $L < {\cal R} < \ell$:
At a fixed distance ${\cal R}$ within this range, the shock wave
is {\it composite}: all intermediate gravitons with
masses\footnote{The extra helicity states of massive gravitons
which plague flat space perturbation theory \cite{vdvz} could be
circumvented in curved backgrounds \cite{nonflat}. For discussions
including the ghosts see \cite{Kogan,kkr,nimags,higuchi}. } $m <
2/{\cal R}$ contribute to it, but only at ${\cal O}(L^2/\ell^2)$
relative to the ultralight mode because of the coupling
suppression. The solution approximates very closely the $4D$
Aichelburg-Sexl solution, $f_{A-S} \simeq \frac{p}{\pi } \,
G_{eff} \, \log \frac{\cal R}{2\ell}$, where the logarithm
gradually changes towards the $AdS_4$ shock wave as distance
increases. Defining $G_{eff}$ as the $4D$ effective gravitational
coupling which controls the deflection imparted by the shock on a
test particle as in $4D$ \cite{gerard,acvms}, we note that it {\it
runs}\footnote{We thank John Terning for related discussions about
such phenomena.}. On the bulk side this is a completely classical
effect. The intermediate gravitons drop out from the resonance as
the distance increases due to Yukawa suppression; after each step
$\sim \ell/n$ of distance, $G_{eff}$ drops by $\frac{\Delta
G_{eff}}{G_{eff}} \sim n \frac{L^2}{2\ell^2}$. These jumps are
sharp, and are the largest at the shortest distances, never
exceeding $L/\ell \ll 1$ in this regime. So overall gravity
remains $4D$. Except for the coupling suppression of all heavier
modes generated by warping, this regime resembles gravity on a
space with a large compact dimension at distances between the
cutoff and the compactification radius \cite{add}.

\item[{{\bf III}\,:}]
$\ell \la {\cal R} < {\ell}^3/L^2$: All heavier gravitons are now
decoupled, and the behavior of the shock wave at these distances
is completely set by the ultralight mode, which mimics $4D$
gravity with great accuracy. To the leading order, the shock wave
behaves as the $AdS_4$ shock, succumbing to the $AdS_4$ curvature.
Although the heavier gravitons have decoupled, the small graviton
mass feeds the weak running of the $4D$ effective gravitational
coupling with distance. It yields subleading distance-dependent
corrections to the shock, which appear as a multiplicative factor
in the amplitude $\sim \Bigl(1 - {\cal O}(1)\frac{L^2}{\ell^2}
\frac{\cal R}{\ell}\Bigr)$, continuing to weaken gravity.

\item[{{\bf IV}\,:}] $\ell^3/L^2 < {\cal R}$: At such large distances
along the brane, the subleading corrections controlled by the mass
of the ultralight graviton pick up. Due to the graviton mass, the
gravitational force weakens faster. However these mass-induced
corrections open a narrow window into the fifth dimension. The
ultralight graviton mass probes the full $5D$ geometry, making the
decay rate of the ultralight graviton wave function a little {\it
slower} than it would have been if the decay rate were controlled
a by a perturbative graviton mass term in $AdS_4$. Probing this
regime one would discover both the graviton mass and the hidden
extra dimension.

\end{itemize}

From the solution, we extract the formula for the $4D$ Planck mass
in terms of the $5D$ Planck mass and the curvature radii, defined
as the inverse coupling of the ultralight mode. It is\footnote{The
(-) sign cancels another (-) sign hidden in
$\partial_\nu\,P_\nu^{-1}$.}
\be M^2_4 = \frac{2 M^3_5 \ell}{2\,\nu+1}
\frac{[-\partial_\nu\,P_\nu^{-1}\,\left(-\cos(y_0)\right)]}
{P_\nu^{-2}\left(-\cos(y_0)\right)} \Big|_{\nu= \nu_0} \, ,
\label{planckmass} \ee
where $P_\nu^k$ are Legendre functions,
$y_0=\arcsin\left(L/\ell\right)$, and $\nu_0 =
[\sqrt{9+4\ell^2\,m^2}-1]/2$ where $m$ is the ultralight mode
mass. This reproduces exactly $M^2_4 = M^3_5 L$ to the leading
order in $L/\ell \ll 1$, supporting the arguments of
\cite{porrati}. However, there are also higher order corrections
in $L/\ell$, showing that the dual $CFT$ is modified in the far
infra-red. This stems from $AdS/CFT$, which relates classical bulk
dynamics with the dual $CFT$ in the large $N$ limit, including
quantum corrections from planar diagrams \cite{adscft}.

To recap, from the bulk view point low energy localized gravity is
born from a conspiracy between the $AdS_4$ and $AdS_5$ curvatures.
Together they give rise to the ultra-light mode. The brane
curvature separates the rest of the spectrum from it by a mass gap
$\sim \ell^{-1}$. In turn, the bulk curvature pulls all modes with
masses $m \ga 1/\ell$ away from the brane,  suppressing their
coupling relative to the ultralight mode by a power $mL^2/\ell$,
as in RS2 \cite{rs2,tunneling}. Thus the shock wave appears $4D$
to the leading order even at distances ${\cal R} < \ell$, at which
many intermediate gravitons contribute to it. Without the
suppression of the couplings the running of $G_{eff}$ would have
been much faster, spoiling the $4D$ guise of the shock wave. At
very large distances, ${\cal R} > \ell$, brane curvature kicks in
before the ultralight mode Yukawa suppression develops. The
ultralight graviton mass still feeds the running of $G_{eff}$,
yielding large distance decay as\footnote{This formula is a
dimensional hybrid: the factor $2$ comes from $AdS_4$, and $1/3$
from $AdS_5$ dynamics. The latter is the culprit of mimicking the
``running" of $G_N$ we mention above. See below.} $\exp[-(2 +
\ell^2 m^2/3) r/\ell]$. This postpones the deviations from $4D$ to
${\cal R} > \ell/(\ell^2 m^2) \sim \ell^3/L^2$.

Our findings are consistent with the interpretation of locally
localized gravity in terms of a defect $CFT$ proposed in
\cite{kara2,defect}. In the picture where the defect is ``fat" the
$CFT$ is cut off in the UV at its inverse thickness, $1/L$. When
the cutoff is fixed to some finite value, the local geometry
becomes dynamical, described by $4D$ gravity whose Planck mass is
set by the cutoff. The defect excitations are described by an
effective conformal field theory emerging from the dual $CFT$
mixing with the defect, but whose conformal symmetry breaks to
$SO(3,2)$ at a low scale $1/\ell$, generated by the strong
coupling dynamics of the system. These degrees of freedom are
light and interact with the $4D$ graviton giving it a small,
radiatively generated mass $m^2 \sim g_* \frac{1}{M^2_4}
\frac{1}{\ell^4} \sim \frac{L^2}{\ell^4}$ \cite{porrati}. We find
that they also must renormalize the $4D$ Newton's constant by IR
contributions, which correspond to the terms in the $L/\ell$
expansion of our formula (\ref{planckmass}): $M_4^2 = M^3_5 L
\Bigl(1 +\frac{5}{12} (\frac{L}{\ell})^2 + \ldots \Bigr)$. They
also dress up the $4D$ Newton's constant with external
momentum-dependent terms, which make the graviton look composite
at distances below $\ell$. Far from the source the effective
Newton's constant goes to zero as distance increases at a rate set
by the anomalous dimension of the operator which should be dual to
the graviton in $AdS_4 \subset AdS_5$. This suggests a link with
the holographic rescaling in the boundary $CFT$ \cite{lenny}.

The paper is organized as follows. In the next section we develop
the formalism needed to the construct the shock waves on $AdS_4
\subset AdS_5$. In section 3 we derive the shock wave profile in a
closed form. We explore the limits of the solution in section 4,
and derive the Planck mass formula (\ref{planckmass}) and explore
the four relevant dynamical regimes, outlining the limit $\ell
\rightarrow \infty$ which takes the theory to RS2. We discuss the
connection with $AdS/CFT$ in section 5 and summarize in section 6.
In the appendices we give some details of the derivation of the
wave profile field equation, the limiting cases of the solution
and the spectrum of localized gravitons.

\section{Setup}

Generalizations of the Aichelburg-Sexl waves, describing a
relativistic particle in a $4D$ flat space \cite{pirani,aichsexl},
to curved spaces and to more dimensions, as well as the properties
of the solutions, have been considered in
\cite{thooft,gabriele,devega,barrabes,hota,pogri,kostas,horit,roberto,kallet,kostnew}.
As in those cases, we find that the gravitational field equations
reduce to a single {\it linear} partial differential equation for
the shock wave. We follow the cutting-and-pasting trick of Dray
and 't Hooft \cite{thooft}, adopted to de Sitter and $AdS$
backgrounds by Sfetsos \cite{kostas}, and recently employed by one
of us in the construction of exact shock waves \cite{kallet} in
DGP braneworlds \cite{dgp}. Many of the results of \cite{kallet}
carry over with minor changes. That is, we start with a background
determining a tensional 3-brane residing in $AdS_5$ bulk such that
the induced geometry on the brane is $AdS_4$. We perturb it with a
photon, which moves along a null geodesic with a momentum $p =
2\pi \nu$. Its momentum generates a gravitational field, since it
contributes to the total stress-energy tensor. Because of the
infinite boost, the gravitational field of the particle will be
completely confined to the instantaneous plane orthogonal to the
direction of motion, and the Einstein's equations will break up
into the background equations and a single field equation for the
wave profile. This equation is linear by the analyticity of the
setup. This can be seen from imagining a solution for a massive
source and expanding the exact metric in a Taylor series in the
mass $m$, and boosting it to relativistic speeds, by enforcing
$\cosh \gamma \rightarrow \infty$, $m \rightarrow 0$ and $m \cosh
\gamma = p = {\rm const}$. Only the linear terms in $m$ in the
expansion survive in this limit because there is only one factor
of the boost parameter $\cosh \gamma$ in the metric, which will be
overcompensated by higher powers of $m$. The linearized solution
becomes exact, implying that the only nontrivial field equation
must be linear. If we choose coordinates such that $v$ runs along
the photon null geodesic and $u$ orthogonally to it, such that the
world-line of the particle is $u=0$, then by causality the field
experiences a jump at $u=0$. The gravitational field of the
particle jolts the observer exactly at the moment when the
particle flies by her \cite{thooft}.

\FIGURE{\epsfig{file=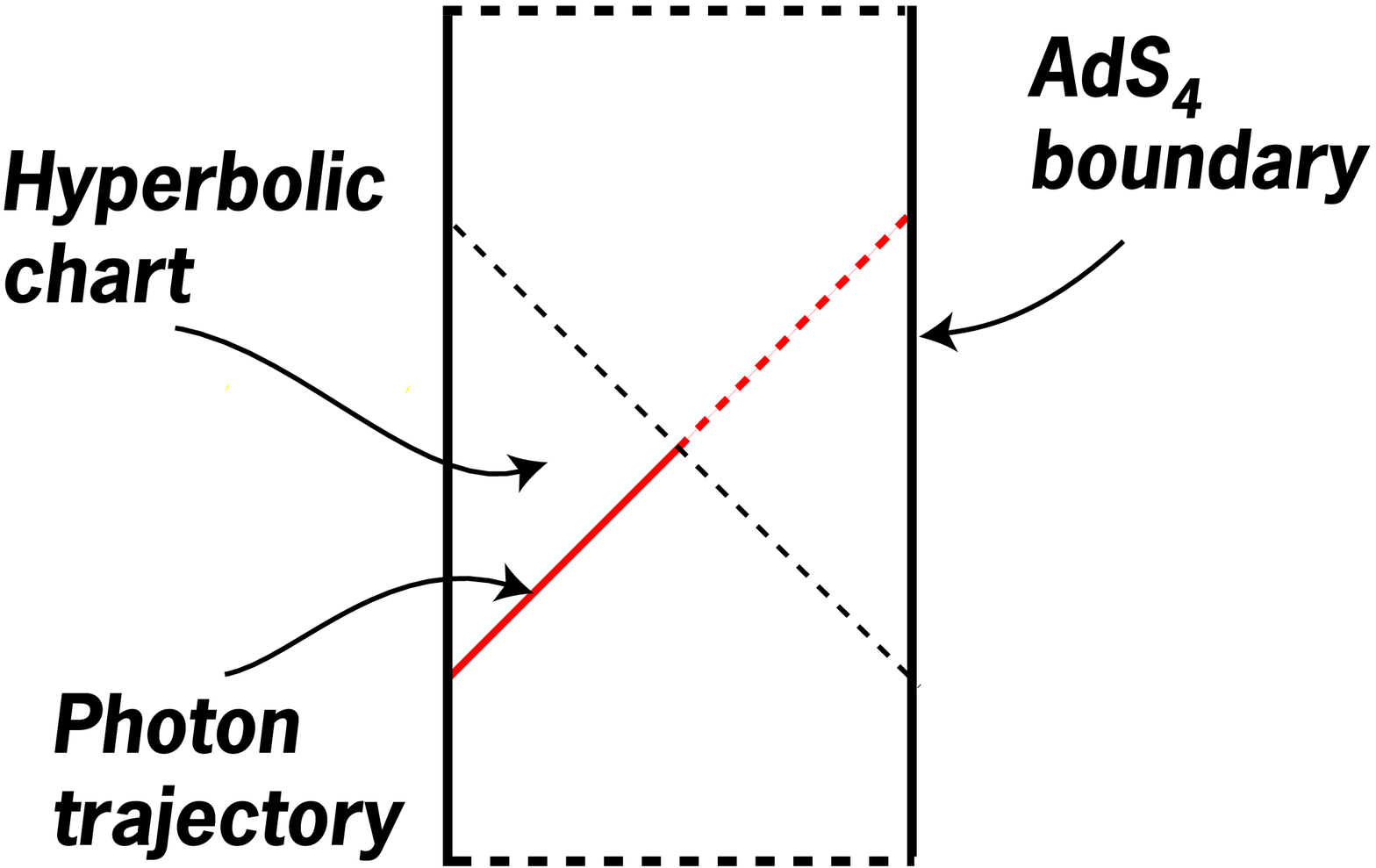, width=11cm, height=7cm}
\caption{$AdS_4$ strip. Patch of $AdS_4$ covered by Eq.
(\ref{branemetric}), designated ``Hyperbolic chart", and the
photon trajectory. The bold red line depicts the part of the
photon trajectory covered by the chart on which the metric
(\ref{branemetric}) is defined, and the dashed red line segment is
its future extension. } \label{figone}}

Our starting point is, as in \cite{kallet}, the $5D$ metric
describing the background, in this case $AdS_4 \subset AdS_5$
\cite{bent,details}, which can be written as a warped product of
$AdS_4$:
\begin{equation}
ds_{AdS_5}^2 = \frac{L^2}{\ell^2 \sin^2 (\frac{|z|+z_0}{\ell})} \,
\Bigl[ ds_{{ {AdS_4}}}^2+ dz^2 \Bigr] \, . \label{metric}
\end{equation}
In this equation the parameter $z_0$, determining the conformal
distance of the brane from the center of the bulk (i.e. the warp
factor ``bounce") is given by $\sin(z_0/\ell)=L/\ell$. This is
just the Israel junction condition in disguise.
Following~\cite{kostas}, we choose a patch of $AdS_4$ covered with
the static coordinates foliated by hyperbolic hyper-planes,
\begin{equation}
ds_{{{AdS_4}}}^2= - \Bigl(\frac{r^2}{\ell^2}-1\Bigr)\, dt^2+
\frac{dr^2}{\frac{r^2}{\ell^2}-1} + r^2\,
\Bigl(d\chi^2+\sinh^2\chi\,d\phi^2 \Bigr) \, . \label{branemetric}
\end{equation}
These coordinates cover the region of $AdS_4$ enclosed by the
$AdS_4$ boundary and the two null cones balancing on each other's
tips, as depicted in Fig. (\ref{figone}). Note that while the
$AdS_4$ boundary is at $r\rightarrow \infty$, the center is at
$r=\ell, |t| \ne \infty$. The surfaces $r=\ell, t \rightarrow \pm
\infty$ are Cauchy horizons. This is the correct $AdS_4$ domain,
bounded by the null geodesics along which we will introduce the
discontinuity \`a la Dray-'t Hooft \cite{kostas} (see also
\cite{pogri} for a discussion of the shape of shocks in $AdS$).
The null coordinates are given by the map
\begin{equation}
u = \ell \,{ e}^{t/\ell}
\left(\frac{r-\ell}{r+\ell}\right)^{1/2}\, , \qquad \qquad v =
\ell\,{ e}^{-t/\ell} \left(\frac{r-\ell}{r+\ell}\right)^{1/2} \, .
\label{coordmaps}
\end{equation}
In terms of these coordinates we can rewrite the brane $AdS$
metric (\ref{branemetric}) as
\begin{equation}
ds^2= \Omega^2(|z|)\, \Bigl[ \frac{4\,du\,dv}{ (1-uv/\ell^2 )^2}
+\ell^2 \, \Bigl(\frac{1+uv/\ell^2}{1-uv/\ell^2} \Bigr)^2 \,
(d\chi^2+\sinh^2\chi\,d\phi^2 )+dz^2\Bigr]\, . \label{nullbackmet}
\end{equation}
We introduce the warp factor $\Omega(|z|) = \Bigl[{L}/[\ell
\sin(\frac{|z|+z_0}{\ell})]\Bigr]$ as a shorthand notation.

Enter the photon. We choose the $v$-axis of (\ref{nullbackmet}) as
its trajectory, i.e. the null geodesic $u=0$. That is depicted as
the red null line in Fig. (\ref{figone}). Then following Dray and
't Hooft we change the metric by including a jump in the $v$
coordinate at $u=0$ \cite{thooft}. This encodes the shock wave
profile. To do this, we can replace $v, dv$ in (\ref{nullbackmet})
by
\begin{equation}
v \rightarrow v + \Theta(u) f \, , \qquad \qquad  dv \rightarrow
dv + \Theta(u) df \, , \label{jump}
\end{equation}
where $f$ is the shock wave profile, which is required to solve
the field equations with the photon contribution to the
stress-energy tensor included on the right hand side (RHS). As
before the wave profile $f$ depends only on the spatial
coordinates transverse to the trajectory of the source, which in
our case are the coordinates on ${\cal H}_2$ and the bulk
coordinate $z$. Here $\Theta(u)$ is the Heaviside step function.
It is convenient to slightly change the coordinates to $\hat v = v
+ \Theta(u) f$, in which case $dv \rightarrow d \hat v - \delta(u)
f du$. Dropping carets, we substitute these in
(\ref{nullbackmet}), yielding the metric {Ansatz} which includes
the shock wave profile:
\begin{equation}
ds^2= \Omega^2(|z|)\, \Bigl[ \frac{4\,du\,dv}{ (1-uv/\ell^2 )^2} -
\frac{4 \delta(u) f du^2}{(1- uv/\ell^2)^2} +\ell^2 \,
\Bigl(\frac{1+uv/\ell^2}{1-uv/\ell^2} \Bigr)^2 \,
(d\chi^2+\sinh^2\chi\,d\phi^2 )+dz^2\Bigr]\, . \label{wave}
\end{equation}
Next we substitute Eq. (\ref{wave}) into the field equations
\begin{eqnarray}
G^{A}_{5\,B} &=& \frac{6}{L^2} \, \delta^A{}_B  - 6 \,
\Bigl(\frac{1}{L^2} \, - \frac{1}{\ell^2}\Bigr)^{1/2} \,
\delta^\mu{}_\nu \, \delta^A_\mu \, \delta^\nu_B  \,  \delta(z)
\nonumber \\
&& ~~~~~~~~~ + 2 \, \frac{p}{M_5^3 } \,
\frac{g_{4\,uv}}{\sqrt{g_5}} \, \delta^\mu_v \, \delta^u_\nu \,
\delta^A_\mu \, \delta^\nu_B \, \delta(\chi) \, \delta(\phi) \,
\delta(u) \, \delta(z) \, . \label{fieldeqs}
\end{eqnarray}
We include the brane-localized terms as $\delta(z)$-function
sources on the RHS because the metric (\ref{wave}) is in the
Gaussian-normal form, and $\sqrt{g_4/g_5} = 1/\Omega = 1$ at the
location of the brane $z=0$. We also include the photon
stress-energy contribution, $\propto p$, in the source on the RHS,
and pick the coordinates on ${\cal H}_2$ such that the photon
trajectory is along $\chi=\phi=0$. In the equation, $G^{A}_{5\,B}$
is the bulk Einstein tensor evaluated on (\ref{wave}), indices
$\{A,B\}$ and $\{\mu,\nu\}$ are bulk and brane worldvolume
indices, respectively, and $g_{4\,\mu\nu}$ is the induced metric
on the brane, in this case simply the obvious restriction of
$g_{AB}$ to $g_{\mu\nu}$. We have traded the brane tension
$\lambda$ from the equation using the relation between it and the
bulk and brane curvature radii \cite{bent}, $\lambda = 6 M_5^3
\Bigl(\frac{1}{L^2} - \frac{1}{\ell^2}\Bigr)^{1/2}$. One can
easily check that the photon stress-energy is automatically
conserved, $\nabla_\mu T^\mu{}_{\nu \, {\rm photon} } \propto u
\delta(u) = 0$, as it must be by Bianchi identities. Relegating
the details to the Appendix A, here we merely quote the result of
this computation: the {\it linear} field equation for the wave
profile is
\begin{equation}
\partial^2_z f +
3\frac{\partial_{|z|}\Omega}{\Omega}\,\partial_{|z|}f+
\frac{1}{\ell^2} \,( \Delta_2 f -2 f) = \frac{2p}{M^3_5 \ell^2}
\,\delta\left(\cosh\chi-1\right)
\,\delta\left(\phi\right)\,\delta\left(z\right) \label{feqn}
\end{equation}
where the operator $\Delta_2$ is the Laplacian on the ``unit"
hyperbolic surface ${\cal H}_2$. This equation arises from the
$u-v$ component of (\ref{fieldeqs}), whereas all others are simple
identities by virtue of the relation $L/\ell = \sin(z_0/\ell)$.
Now we can turn to solving (\ref{feqn}).

\section{Shocks}

The wave profile can only depend on the transverse spacelike
coordinates $z, \chi, \phi$. Having picked the transverse
coordinates such that the photon trajectory is along $\chi=\phi=0$
in the ${\cal H}_2$ hyperplane, we need only look at the bottom
portion of the trajectory, starting at the $AdS_4$ boundary and
ending at the $AdS_4$ center. These coordinates provide a natural
description of the shock wave profiles in $AdS_4$ because the
planes $t,r= const$ are hyperbolic, exactly the shape of the shock
wave surfaces \cite{pogri,kostas}. This is the part in the past of
the chart covered by the coordinates featuring in the metric
(\ref{nullbackmet}), as depicted by the bold red line in Figure
(\ref{figone}). Indeed, since the photon trajectory is $u=0$, from
(\ref{coordmaps}) we see that it zips along $r= \infty,
t\rightarrow -\infty$ from the $AdS_4$ boundary to its center. We
can easily extend the trajectory beyond the $AdS_4$ center,
mapping it to the future portion by a global $AdS_4$ rotation and
a time reversal. Then, from the axial symmetry around the photon
trajectory, we expect that the wave profile should be independent
of $\phi$, thanks to our choice of coordinate system in the
transverse ${\cal H}_2$. This is all borne out by the solution, as
we will now show.

In the expanded form, using $\Delta_2 = \partial_\chi^2+\coth\chi
\,\partial_\chi +\frac{\partial_\phi^2}{\sinh^2\chi} $ the wave
profile equation (\ref{feqn}) reads
\begin{eqnarray}
\partial^2_y f-3\,\cot(|y|+y_0)\,\partial_{|y|} f +
\Bigl(\partial_\chi^2 f +\coth\chi \,\partial_\chi f
+\frac{\partial_\phi^2 f}{\sinh^2\chi}-2 f \Bigr) = ~~~~~~~~ &&
\nonumber
\\
= \frac{2\,p}{M^3_5\ell}\,\delta\left(\cosh\chi-1\right)
\,\delta\left(\phi\right)\,\delta(y) \, , && \label{fineq}
\end{eqnarray}
where we introduce $y=z/\ell$,
$y_0=z_0/\ell=\arcsin\left(L/\ell\right)$ to simplify the
subsequent manipulations.

This equation is separable. We therefore seek for solutions of the
form
\begin{equation}
f\left(y,\,\chi,\,\phi\right)=\sum_m\,\int dq\,
\psi_{q,m}\left(y\right)\, H_{q,m}\left(\chi,\,\phi\right)\,.
\end{equation}
The functions $H_{q,m}\left(\chi,\,\phi\right)$ are the
eigenfunctions of the Laplacian on ${\cal H}_2$,
\begin{equation}
\Delta_2 H_{q,m}\left(\chi,\,\phi\right)=-\left(q^2
+\frac{1}{4}\right)H_{q,m}\left(\chi,\,\phi\right)\,.
\end{equation}
They are defined by \cite{bander,grosche}
\begin{equation}
H_{q,m}\left(\chi,\,\phi\right)={e}^{im\phi}
\,Z_{q,m}\left(\chi\right)\, , \qquad
Z_{q,m}\left(\chi\right)\equiv
\frac{\Gamma\left(iq+m+1/2\right)}{\Gamma\left(iq\right)}
\,P^{-m}_{iq-1/2}\left(\cosh\chi\right)\, .
\end{equation}
They are an orthonormal basis on ${\cal H}_2$ by $\int_0^\infty
d\chi\, \sinh\chi\, Z_{q,m}\left(\chi\right)\,
Z^*_{q',m}\left(\chi\right)=\delta\left(q-q'\right)$, $ \int
d\phi\,{e}^{im\phi}\,{e}^{-im'\phi}=2\pi\,\delta_{m,m'}$. Note
that $H_{q,m}\left(\chi,\,\phi\right)$ are the generalization of
spherical harmonics $Y_{q,m}\left(\theta,\,\phi\right)$ to the
hyperplane ${\cal H}_2$.  Their completeness relation
\cite{grosche} with our choice of coordinate system on ${\cal
H}_2$ yields
\begin{equation}
\sum_{m=-\infty}^\infty\int_0^\infty
dq\,H_{q,m}\left(\chi,\,\phi\right)\,
H^*_{q,m}\left(0,\,0\right)=\delta\left(\cosh\chi-1\right)
\,\,2\pi\,\delta\left(\phi\right).
\end{equation}
Inserting this for $\delta\left(\cosh\chi-1\right)
\delta\left(\phi\right)$ in (\ref{fineq}) and using the
orthonormality of the eigenmodes, we get the equation for the
radial bulk modes $\psi_{q,m}$. The Fourier coefficients of the
$\delta\left(\cosh\chi-1\right) \delta\left(\phi\right)$ in this
basis obey $Z^*_{q,0}\left(0\right)=
\Gamma\left(-iq+1/2\right)/\Gamma\left(-iq\right)$, $Z^*_{q,m\neq
0}\left(0\right)=0$, and so only the axially symmetric modes,
$m=0$, will have a nonzero source. Since we are only interested in
the field of the photon, carried by its momentum $p$, and not in
the homogeneous modes, we set $\psi_{q,m\ne 0} = 0$ for all $q$.
Hence the equation for the axially symmetric modes,
$\psi_{q}\left(y\right)\equiv \psi_{q,0}\left(y\right)$, is
\begin{equation}
\partial_y^2\psi_{q}\left(y\right)-3\,
\cot(|y|+y_0) \,\partial_{|y|}
\psi_q\left(y\right)-\left(q^2+\frac{9}{4}\right)\,
\psi_q\left(y\right)=\frac{p}{\pi M^3_5 \ell}\,
\frac{\Gamma\left(-iq+1/2\right)}{\Gamma\left(-iq\right)}\,
\delta\left(y\right) \label{locgrav}
\end{equation}
One linearly independent set of eigenmodes of the operator on the
LHS is given by the Legendre functions
\begin{equation}
\sin^2\left(|y|+y_0\right)\,
P_{iq-1/2}^{-2}\left(\cos\left(|y|+y_0\right)\right)\,,\,\,\,\,
\sin^2\left(|y|+y_0\right)\,
Q_{iq-1/2}^{-2}\left(\cos\left(|y|+y_0\right)\right)\,\,.
\end{equation}
Notice that in spite of the appearances, these functions are real.
With our coordinate choices, the $AdS_5$ boundary lies at
$|y_b|+y_0=\pi$. Thus
locally localized gravity corresponds to the
linear combination of these two modes which remains regular as the
argument of $P_\nu^{-2}$ and $Q_\nu^{-2}$ approaches $-1$. To find
that combination, we expand the modes in the limit
$\cos\left(|y|+y_0\right) =x \rightarrow -1^+$, where
\begin{equation}
P^{-2}_\nu\left(x\right)= \frac{2}{\Gamma(3+\nu)\,\Gamma(2-\nu)}
\, \frac{1}{1+x} \,, \qquad Q^{-2}_\nu\left(x\right) = -
\frac{\cos\left(\pi\nu\right) \,\Gamma(\nu-1)}{\Gamma(\nu+3)} \,
\frac{1}{1+x}\, . \label{legexpansion}
\end{equation}
The linear combination which is regular at the $AdS_5$ boundary,
and so describes the localized graviton modes, is proportional to
\cite{bateman}
\begin{equation}
\cos\left(\pi\nu\right)\,P_\nu^{-2}
\left(\cos\left(|y|+y_0\right)\right)+2\,\frac{Q_\nu^{-2}
\left(\cos\left(|y|+y_0\right)\right)}{\Gamma(\nu-1)\,\Gamma(2-\nu)}
=P_\nu^{-2}\left(-\cos\left(|y|+y_0\right)\right)\,\,.
\end{equation}
Therefore, the localized graviton modes are
\begin{equation}
\psi_q\left(y\right)\equiv
N_q\,\sin^2\left(|y|+y_0\right)\,P_{iq-1/2}^{-2}
\left(-\cos\left(|y|+y_0\right)\right)\, . \label{modes}
\end{equation}
The normalization constant $N_q$ is determined by a pillbox
integration of the equation (\ref{locgrav}), imposing the $Z_2$
orbifold boundary conditions on the brane at $y=0$. This gives
\begin{equation}
\psi_q'\left(0\right)=\frac{p}{2 \pi M^3_5 \ell}\,
\frac{\Gamma\left(-iq+1/2\right)}{\Gamma\left(-iq\right)}\,,
\label{boundcond}
\end{equation}
and, using the identity
$\left(\sin^2\xi\,P_\nu^{-2}\left(-\cos\xi\right)\right)'
=-\sin^2\xi\,P_\nu^{-1}\left(-\cos\xi\right)$ we finally obtain
\begin{equation}
N_q=-\frac{p\,\ell}{2\pi M^3_5 L^2}\,
\frac{\Gamma\left(-iq+1/2\right)}{\Gamma\left(-iq\right)}
\,\left[P^{-1}_{iq-1/2}\left(-\cos(y_0)\right)\right]^{-1} \, .
\end{equation}

Therefore our complete solution describing the shock wave profile
is given by the integral formula
\begin{eqnarray}\label{master5d}
f\left(y,\,\chi\right) &=& -\frac{p\,\ell}{2\pi M^3_5
L^2}\int_0^\infty dq\,q\tanh\left(\pi\,q\right) \times \nonumber
\\
&& ~~~~~ \times  \sin^2\left(|y|+y_0\right)
\frac{P_{iq-1/2}^{-2}\left(-\cos\left(|y|+y_0\right)\right)}
{P_{iq-1/2}^{-1}\left(-\cos\left(y_0\right)\right)}
P_{iq-1/2}\left(\cosh\chi\right)\, ,
\end{eqnarray}
where we have used $\left|\Gamma\left(iq+1/2\right)/
\Gamma\left(iq\right)\right|^2=q\,\tanh\pi q$. On the brane, at
$y=0$, the integral formula (\ref{master5d}) reduces to
\begin{equation}\label{master}
f\left(0,\,\chi\right)=-\frac{p}{2\pi M^3_5 \ell}\int_0^\infty
dq\,q\tanh\left(\pi\,q\right)
\frac{P_{iq-1/2}^{-2}\left(-\cos(y_0)\right)}
{P_{iq-1/2}^{-1}\left(-\cos(y_0)\right)}
P_{iq-1/2}\left(\cosh\chi\right) \, .
\end{equation}
These two equations represent our main technical results. We now
turn to extracting physics from them.

\section{Limits}

Using the integral formula (\ref{master}) we can now probe the
shock wave profile along the brane as a function of the transverse
distance from the photon which sources the field. We compare the
result in different regimes with the form of the shock wave in
conventional $GR$ in $5D$ and $4D$ to elucidate the nature of
localized gravity. Note that in general this approach could be
suspect because of the issues of gauge choice, related to the
choice of the background metric. However because the relativistic
source excites only the transverse traceless gravitons, and the
gravitational dynamics linearizes in this limit, no gauge
ambiguities arise in any coordinate cover of $AdS_4 \subset
AdS_5$. The transverse traceless nature of the shock ensures that
we are working in the unitary gauge and that information extracted
from the shock wave describes the physical gravitons of the
theory, independently of the background coordinates. Simply put,
the shock wave amplitude which we compute is the static limit of
the physical Green's function of the tensor graviton resonance in
the hyperplane transverse to the motion of the relativistic source
\cite{gerard,acvms}. The discussion below will elaborate this
further.

We still need to define a measure of the transverse distance
${\cal R}$ from the source in the background metric (\ref{wave}).
We note that the metric transverse to the photon on the brane is
$ds_2^2|_{z=u=0} = \ell^2 \, (d\chi^2+\sinh^2\chi\,d\phi^2 )$.
Thanks to the axial symmetry of (\ref{metric}) we need only
consider variation of $f$ with the ``radial" transverse distance
on ${\cal H}_2$. Thus we should scan the wave variation with
$\chi$. The proper radial distance from the source is therefore
measured by ${\cal R} = \ell \chi$. At very large distances $\chi
\gg 1$ where we can't neglect the $AdS_4$ curvature, the warping
of $AdS_4$ changes the measure of angular distances to the
exponent $\exp(\chi) = \exp({\cal R}/\ell)$, but we needn't pay a
particular attention to it here thanks to the axial symmetry of
the shock waves.

\subsection{Surfing the Wave}

Now, to test locally localized gravity we need to define the
benchmarks to compare our shock wave solution (\ref{master}) to.
At short distances, we should compare it to the $5D$ version of
the Aichelburg-Sexl wave, derived in \cite{gabriele,devega}
\be f_{5D}({\cal R}) = - \,
\frac{p}{2\,\pi\,M^3_5}\,\frac{1}{{\cal R}}  \, . \label{5dshock}
\ee
At large distances ${\cal R} > L$, the most convenient form of the
shock wave in $AdS_4$ for our purposes is the one derived by
Sfetsos \cite{kostas},
\begin{equation}
f_{4D ~AdS}(\chi) = - \, \frac{p}{\pi M^2_4}  \,
Q_{1}\left(\cosh\chi\right) \, . \label{4dshock}
\end{equation}
This solution describes the $4D$ $GR$ shock wave at all length
scales in $AdS_4$, and correctly reproduces the $4D$
Aichelburg-Sexl wave in flat space at short distances ${\cal R}
\ll \ell$, where $AdS_4$ curvature is negligible\footnote{While
the constant in (\ref{aiseshock}) is gauge-dependent in flat
space, the gauge is uniquely fixed by the choice of the
normalization of ${\cal R}$ under the logarithm \cite{thooft}. We
will use it later, when discussing the limit to RS2.}:
\begin{equation}
f_{4D\, {\rm A-S}} =  \frac{p}{\pi\,M_4^2}\,\log\frac{\cal R}{2
\ell} \, + \, \frac{p}{\pi\,M_4^2} \, + \, \ldots  \,  .
\label{aiseshock}
\end{equation}
Both solutions (\ref{4dshock}), (\ref{aiseshock}) will provide
useful standards for our ends. Before we continue with the
comparisons it is useful to review the salient features of the
spectrum of locally localized gravity.

\subsubsection{Spectral Decomposition of the Shock Wave}

Let us briefly review some useful mathematical tools before
exploring the physics of the shock wave. First, it is convenient
to re-express the integral on the RHS of (\ref{master}) as a sum.
The series representation of (\ref{master}) is (for details see
Appendix B):
\begin{equation}
f\left(0,\,\chi\right)=\frac{p}{2\,\pi\,M^3_5\,\ell}
\,\sum_{\nu>-1/2}\left(2\,\nu+1\right)\,\frac{P_\nu^{-2}\left(-\cos(y_0)
\right)}{\partial_\nu\,P_\nu^{-1}\,\left(-\cos
(y_0)\right)}\,Q_\nu\left(\cosh\chi\right)\, . \label{shockseries}
\end{equation}
The sum runs over all the values of $\nu \equiv iq-1/2$ that solve
\be P_\nu^{-1}\,\left(-\cos(y_0)\right)=0 \, , \label{eignens} \ee
with the condition $\nu>-1/2$. This is an exact statement, no
approximations have been made in writing (\ref{shockseries}),
(\ref{eignens}).

The equation (\ref{shockseries}) is the graviton mode expansion of
the shock wave profile. We note that the expression
(\ref{eignens}) {\it is} the secular equation of the problem,
determining the spectrum of gravitons in locally localized
gravity.  This can be verified easily from the homogeneous
equation of (\ref{locgrav}), which is
\begin{equation}
\partial_y^2\psi\left(y\right)-3\,
\cot(|y|+y_0) \,\partial_{|y|}
\psi\left(y\right)=m^2\ell^2\,\psi\left(y\right)\, ,
\label{modeeq}
\end{equation}
after substituting $m^2 = \frac{\nu\left(\nu+1\right)-2}{\ell^2}$.
Upon setting $p=0$, the boundary condition (\ref{boundcond})
reduces to the Neumann boundary condition
\begin{equation}
\psi_q'\left(0\right)= 0 \,. \label{neumann}
\end{equation}
Plugging in the localized graviton wave functions (\ref{modes}),
we find that those which satisfy (\ref{neumann}) are counted by
the index $q$, defined such that $\nu =  iq-1/2$ solves
(\ref{eignens}) for a fixed $y_0$. They yield the mass spectrum
through the relation of $\nu$ and the graviton mass $m$. The
graviton couplings are given by the coefficients of
$Q_\nu\left(\cosh\chi\right)$ in the series (\ref{shockseries}):
\be g^2_n = \frac{2\,\nu+1}{2 M^3_5 \ell}
\frac{P_\nu^{-2}\left(-\cos(y_0)\right)} {[ -\,
\partial_\nu\,P_\nu^{-1}\,\left(-\cos(y_0)\right)]} \Big|_{\nu=
\nu_n} \, . \label{coupl} \ee
The (-) sign cancels another (-) sign concealed in
$\partial_\nu\,P_\nu^{-1}$. This is the exact formula for the
couplings of each individual graviton mode to the matter on the
brane in terms of the $5D$ Planck mass and the curvature radii $L$
and $\ell$.

We can find the explicit values of the masses and couplings in the
limit $L \ll \ell$ (see Appendix C for the derivation). For
ultralight and intermediate modes, $\nu_n \simeq
n+\frac{L^2}{4\ell^2}\,n(n+1)$ (\ref{nukpiccolo}), whereas for
heavy modes $\nu_n \simeq n\left(1+ \frac{L^2}{\pi \ell^2}
\right)$ (\ref{largeeig}). The low end masses are
\be m^2 = \frac{1}{\ell^2} \Bigl[(n-1)(n+2) + \frac{1}{2}
n(n+1/2)(n+1) \frac{L^2}{\ell^2} \Bigr] \, + \ldots \,, \qquad n =
1, 2, 3, \ldots \, .\label{eigenmasses} \ee
The ultralight mode of mass $m = \sqrt{\frac32}\frac{L}{\ell^2} +
\ldots $ appears at $n=1$, as in \cite{misa}. The masses of
intermediate modes are $m \sim \sqrt{(n-1)(n+2)}/\ell$. The masses
of heavy modes are (see (\ref{largeeig}))
\be m^2 = \frac{1}{\ell^2} \Bigl[ n^2 + n + \frac{2L^2}{\pi
\ell^2} n^2 \Bigr] \, + \ldots \, , \qquad n \ga \frac{\ell}{L}
.\label{eigenmassh} \ee

The heavy modes reside above the UV cutoff $1/L$, $m \ga 1/L$, of
the dual $CFT$, where the $CFT$ breaks down, but are still well
defined on the bulk side as long as $L \ll 1/M_5$. They are
effectively decoupled at all distances ${\cal R} > L$. The
couplings of light and intermediate modes are (see Appendix C)
\be g^2_n = \frac{L}{4 M^3_5 \ell^2} \, \frac{(2n+1)n(n+1)}{n+2}
\, \frac1{n-1+n(n+1)(L^2/4\ell^2)} \, + \, \ldots \, , \qquad n
\ge 1 \, . \label{modecoupls} \ee
The ultralight mode is {\it much} more strongly coupled than the
rest, because
\be g^2_1 = \frac{1}{M^3_5 L}  \, + \, \ldots \, ,
\label{modezerocoupls} \ee
just as in RS2, while the couplings of heavier modes are
suppressed relative to it by two extra powers of $L/\ell$:
\be g^2_m \simeq  \frac{m L}{2M^3_5 \ell} \, + \, \ldots \, .
\label{modemasscoups} \ee
This formula eventually saturates to $g^2_m \simeq \frac{1}{M^3_5
\ell}$ at the high end of the spectrum.

We will also need the short and long distance behavior of
$Q_{\nu_n}$'s in order to interpret the physical meaning of the
series (\ref{shockseries}). Short distance limit corresponds to
$\chi\ll 1$, in which case we can use the expansion \cite{bateman}
\begin{equation}\label{apprq}
Q_{\nu_n}\left(\cosh\chi\right)\simeq
-\log\chi/2-\psi\left(\nu_n+1\right) -\gamma_E + {\cal O}(\chi^2)
+ {\cal O}(\chi^2 \ln \chi) + \ldots \, ,
\end{equation}
where $\gamma_E$ is Euler-Mascheroni constant, and
$\psi\left(\nu_n+1\right)$ the digamma function \cite{gradrhyz}.
It is clear that as $\chi$ increases, the logarithm decreases, and
at some point it stops to dominate. Because the $\psi$ function is
approximated by $\psi\left(\nu_n+1\right)\simeq \log\nu_n$ for
large values of $\nu_n$, the expansion (\ref{apprq}) is valid when
the first term dominates over the second, $\nu_n < 2/\chi$. Since
$m \sim \nu_n/\ell$ for intermediate and massive modes
(\ref{eigenmasses}), (\ref{eigenmassh}),  the approximation of
their mode functions (\ref{apprq}) is valid to distances
${\cal R} < 2/m$. On the other hand, for the ultralight mode its
mass $m \sim L/\ell^2$ does not affect the mode function until
much farther out, but the approximation (\ref{apprq}) still breaks
down at ${\cal R} \la \ell$ even for it.

To understand their behavior at very large transverse distances
from the source, we need a different asymptotic formula for
functions $Q_{\nu_n}(\cosh \chi)$. At $\chi \gg 1$ the leading
order behavior is \cite{bateman}
\be Q_{\nu_n}(\cosh \chi) = {\sqrt{\pi}} \, \frac{\Gamma\left(1+
\nu_n \right)}{\Gamma\left(3/2+\nu_n\right)}\,{e}^{-(1+\nu_n)\chi}
\, + \ldots \, . \label{qchilarge}\ee
The scaling $\sim \exp[-(1+\nu_n)\chi]$ is dictated by the mass
and the $AdS$ geometry, as is seen from the mode equation on
${\cal H}_2$, that follows from (\ref{fineq})  by setting the RHS
to zero, separating out the bulk dependence by using
(\ref{modeeq}) and substituting $2+m^2\ell^2 = \nu_n(\nu_n+1)$.
The resulting mode equation on ${\cal H}_2$ is
\begin{equation}
\left(\partial_\chi^2+\frac{\cosh\chi}{\sinh\chi}\,\partial_\chi
-\nu_n(\nu_n+1)\right) \,Q = 0 \,. \label{qdeq}
\end{equation}
It's easy to check that the asymptotic limit (\ref{qchilarge}) is
the large $\chi$ limit of the solution which is regular at
infinity. Rewriting (\ref{qchilarge}) in terms of the mode mass by
using $\nu_n = (\sqrt{9+4 m^2 \ell^2} - 1)/2$ for localized
solutions we get
\be Q_{\nu_n}(\cosh \chi) = {\sqrt{\pi}} \,
\frac{\Gamma\left(\frac{\sqrt{9+4 m^2 \ell^2} + 1}{2}
\right)}{\Gamma\left(\frac{\sqrt{9+4 m^2 \ell^2} + 2}{2}\right)}\,
\exp(-\frac{\sqrt{9+4 m^2 \ell^2} + 1}{2}\chi ) \,  + \ldots \, .
\label{qchilargem}\ee
The intermediate and heavy mode masses are $m^2 \ell^2 > 1$, and
so they will drop off as $\propto \exp[-(m\ell+\frac12) \chi]$,
feeling conventional Yukawa suppressions. This is obvious for the
modes with masses $m \ell \ga {\cal O}(10)$, but even for the few
lighter ones it's clear that Yukawa term cannot be ignored.

On the other hand, for the ultralight mode $m^2 \ell^2 \sim
(L/\ell)^2 \ll 1$, and so at large distances, the ultralight mode
function scales to the leading order as $\propto \exp(-2 \chi)$.
What that means is, the ultralight mode function at large
distances is predominantly controlled by the $AdS_4$ geometry, in
qualitatively the same way light bulk modes in $AdS/CFT$. Indeed,
for the ultralight graviton we can rewrite (\ref{qchilargem})
approximately as
\be Q_{\nu_n}(\cosh \chi) = \frac43 \, e^{- (2+ \frac{m^2
\ell^2}{3})\chi} \,  + \ldots \, . \label{qchilargezero}\ee
Using ${\cal R} = \ell \chi$, this means that the mass remains
negligible all the way out to distances ${\cal R} \ga \ell/(m^2
\ell^2) \simeq \ell^3/L^2$, when it starts slowly ``renormalizing"
the coefficient $4/3$ of $e^{- 2\chi}$. So unlike for the heavier
modes, the regimes where the ultralight mode is well described by
(\ref{apprq}) and by (\ref{qchilarge}) {\it do not} overlap.
Instead, the function $Q_{\nu_1}$ is extremely well approximated
by $Q_1$ all the way out to $\ell^3/L^2$. Because $\nu_1 = 1 +
\frac{L^2}{2\ell^2} + \ldots$, writing
\begin{equation}
Q_{\nu_1}\left(\cosh\chi\right) = Q_{1}\left(\cosh\chi\right)\, +
\, \frac{L^2}{2\ell^2} \,
\partial_\nu Q_{\nu}\left(\cosh\chi\right)
\Big|_{\nu=1} + \ldots \, , \label{qone}
\end{equation}
we find from (\ref{qchilarge}) that since $\partial_\nu
Q_{\nu}\left(\cosh\chi\right) \simeq \chi
Q_{\nu}\left(\cosh\chi\right)$ the corrections to $Q_1$ indeed do
remain small all the way out to $\frac{L^2}{\ell^2} \chi \ga 1$,
or therefore ${\cal R} \ga \ell^3/L^2$.

To summarize: we see that for $n > 1$, the key features of
$Q_{\nu_n}$'s are captured by the representation
\be Q_{\nu_n} \simeq \cases{ -\log\frac{\cal R}{2 \ell}
-\psi\left(m\ell+\frac12\right) -\gamma_E \, , & {\rm for} ~${\cal
R} <2/m $ \, ; \cr {\sqrt{\pi}} \, \frac{\Gamma\left(m\ell+\frac12
\right)}{\Gamma\left(m\ell+1\right)}\,
e^{-(m\ell+\frac12)\frac{R}{\ell}} \, , & ~~~~\, ${\cal R} > 2/m $
\, .} \label{asympqmass}\ee
The two limits practically overlap: at distances ${\cal R} \sim
2/m$, $Q_{\nu_n}$ sharply changes from a logarithm of the distance
to a Yukawa-decaying exponential. On the other hand, the
ultralight mode asymptotics is accurately captured by
\be Q_{\nu_1}  \simeq \cases{ Q_1\left(\cosh(\frac{\cal
R}{\ell})\right) \, , & {\rm for} ~${\cal R} < \ell^3/L^2 $ \, ;
\cr \frac43 \, e^{- (2+ \frac{m^2 \ell^2}{3}) \frac{\cal R}{\ell}
} \, , & ~~~~\, ${\cal R} > \ell^3/L^2 $ \, ,}
\label{asympqzero}\ee
never displaying Yukawa suppression.

With this at hand, we have the tools needed to complete the
analysis of (\ref{shockseries}). First off, using (\ref{coupl}) we
can rewrite (\ref{shockseries}) as
\begin{equation}
f\left(0,\,\chi\right)= - \frac{p}{\pi} \, \sum_{n=1}^{\infty} \,
g_n^2 \, Q_{\nu_n}\left(\cosh\chi\right) \,  \, .
\label{shockseriesg}
\end{equation}
We now need to extract $\propto Q_1$ contribution from
(\ref{shockseriesg}) to compare it to the $4D$ forms of the shock
wave (\ref{4dshock}), (\ref{aiseshock}).  The complication is that
at distances ${\cal R} < \ell$ many massive modes contribute to
it, as is clear for example from Eq. (\ref{apprq}), which shows
that each $Q_{\nu_n}\left(\cosh\chi\right)$ contains small tails
$\propto Q_{1}\left(\cosh\chi\right)$. Indeed, the functions
$Q_{\nu_n}\left(\cosh\chi\right)$ are {\it not} orthogonal
\cite{bateman}. However: from the qualitative expressions for
$Q_{\nu_n}$'s in Eqs. (\ref{asympqmass}), (\ref{asympqzero}) we
know that the regions where they overlap with $Q_1$ are narrow,
their width inversely proportional to the mass of the mode.
Moreover, higher mode ($n>1$) contributions are also coupling
suppressed, by Eqs. (\ref{modezerocoupls}), (\ref{modemasscoups}).
Using this we can systematically extract physical information from
the shock wave.

\subsubsection{Short Distance Behavior}

At distances ${\cal R} \ll L$ the shock wave reduces to the $5D$
version of Aichelburg-Sexl solution in flat space
\cite{gabriele,devega}. We have to take the limit $\chi \ll 1$ of
(\ref{master}) cautiously, to avoid spurious singularities that
appear to plague the expression if $\chi$ were set to zero
directly in the integrand. To do it, we integrate (\ref{master})
first and then take the limit $\chi\rightarrow 0$. In this regime,
the integral is dominated by its large $q$ behavior, and so the
leading short distance contributions come from the leading order
$q$-dependence of the integrand. This is, of course, identical to
the short distance limit considered in \cite{kallet}, because
$q/\ell$ measures the momentum transfer between the source and a
probe. At short distances it is large, and so the shock wave is
dominated by the heavy gravitons whose sheer number overwhelms all
other modes. Using \cite{gradrhyz} (see Eqs. (8.721.1) and
(8.723.1)) we can expand
\begin{eqnarray}\label{frlq}
\frac{P_{iq-1/2}^{-2}\left(-\cos(y_0)\right)}
{P_{iq-1/2}^{-1}\left(-\cos(y_0)\right)} &=&  1/q + \ldots \, , \nonumber \\
P_{iq-1/2}\left(\cosh\chi\right) &=& \frac{1}{\sqrt{\pi}}\left[
\frac{\Gamma\left(iq\right)}{\Gamma\left(1/2+iq\right)}\,
\frac{{e}^{\left(iq+1/2\right)\,\chi}}{\sqrt{{e}^{2\chi}-1}}
+{\mathrm {h.c.}}\right] + \ldots \,,
\end{eqnarray}
and, after using asymptotic expansions for $\Gamma$-functions and
rescaling $q = \zeta/\chi$, we finally find the leading order
contribution to be
\begin{equation}
f\left(0,\,\chi\ll 1\right) = -\frac{p}{2\pi M^3_5 \ell \chi}\,
\sqrt{\frac{2}{\pi}}\,\int_0^\infty d \zeta\,
\frac{\sin\left(\zeta \right)}{\sqrt{\zeta}}\, .
\end{equation}
This integral is easy, it is just $\sqrt{\pi/2}$. Using ${\cal R}
= \ell\chi$ we finally get
\begin{equation}
f_{\rm short \, distance} =
-\frac{p}{2\,\pi\,M^3_5}\,\frac{1}{{\cal R}} \, .
\end{equation}
Thus as anticipated we exactly recover the $5D$ Minkowski space
shock wave profile (\ref{5dshock}) of \cite{gabriele,devega}. This
is a nice consistency check of the method and the calculation.

\subsubsection{Composite Graviton Regime}

When ${\cal R}>L$, the best form of our solution to work with is
the mode expansion (\ref{shockseriesg}). To explore this region we
start with ${\cal R} \ll \ell$ and move out. In this regime the
relevant $4D$ benchmark to compare (\ref{shockseriesg}) to is the
limiting form of (\ref{4dshock}) given, to the leading order, by
the $4D$ Aichelburg-Sexl shock wave (\ref{aiseshock}). For a fixed
${\cal R} \ll \ell$, we see from Eq. (\ref{asympqmass}) that all
the gravitons with masses $m > 2/{\cal R}$ are effectively
decoupled because of the Yukawa suppression: their contribution to
(\ref{shockseriesg}) is exponentially suppressed, $\propto \exp(-m
{\cal R})$, over and above the coupling suppressions of
(\ref{modemasscoups}). Thus we can truncate the sum in
(\ref{shockseriesg}) to $m < 2/{\cal R}$, or alternatively to $n
\la 2 {\ell}/{\cal R}$ (which will clearly still involve many
terms for a small ${\cal R}$). The leading contribution of all of
these modes to (\ref{shockseriesg}) will be $\propto \frac{p}{\pi}
g_n^2 \log\frac{\cal R}{2\ell}$, and so we conclude that the
leading term in (\ref{shockseriesg}) will be of the form
\begin{eqnarray}
f_{0} &=&  \frac{p}{\pi}\, \Bigl( g_1^2 + \sum_{n=2}^{m_{max} <
2/{\cal R}} g_n^2 \Bigr) \, \log\frac{R}{2 \ell} \, \nonumber \\
&+& \frac{p}{\pi} \,  \Bigl(  g_1^2 \, [\psi(\nu_1+1) + \gamma_E]
+ \sum_{n=2}^{m_{max} < 2/{\cal R}} g_n^2 \, [\psi(\nu_n+1) +
\gamma_E] \Bigr) \, + \, \ldots \, . \label{fzero}
\end{eqnarray}
Here we also retain subleading corrections to compare with the
limiting case of shock wave in RS2 later on. By Eqs.
(\ref{modezerocoupls}), (\ref{modemasscoups}) we can write $g_n^2
= \frac{L}{2\ell} mL g_1^2 + \ldots$, so that
\begin{eqnarray}
f_{0} &=&  \frac{p}{\pi}\, g_1^2 \,  \Bigl( 1 +  \frac12
\sum_{n=2}^{m_{max} < 2/{\cal R}} \frac{L}{\ell} mL + \ldots
\Bigr) \, \log\frac{R}{2 \ell} \, \nonumber \\
&+& \frac{p}{\pi} \,  \Bigl(  g_1^2 \, [\psi(\nu_1+1) + \gamma_E]
+ \sum_{n=2}^{m_{max} < 2/{\cal R}} g_n^2 \, [\psi(\nu_n+1) +
\gamma_E] \Bigr) \, + \, \ldots \, . \label{fzeroone}
\end{eqnarray}
Comparing (\ref{fzeroone}) with (\ref{aiseshock}), one can
identify the coefficient of the logarithm in (\ref{fzeroone}) with
the $4D$ effective gravitational coupling,
\be G_{eff} = g_1^2 \,  \Bigl( 1 + \frac12 \sum_{n=2}^{m_{max} <
2/{\cal R}} \frac{L}{\ell} mL + \ldots  \Bigr) \, . \label{Geff}
\ee
The logic behind this is that in $4D$ $GR$, the shock wave is the
Fourier transform of the gravitational elastic forward scattering
amplitude between the source and the probe \cite{gerard,acvms}.
Thus if we took (\ref{fzeroone}), Fourier-transformed it to the
momentum picture, and factored out the leading order momentum
dependence characterizing the $4D$ interaction, the remainder
would measure the interaction strength at a given scale. In
quantum field theory such identification yields a coupling which
is {\it scale-dependent} when loops are included: because of
renormalization effects, it depends on the energy scale, or its
conjugate distance scale, at which it is measured. The scale
dependence in (\ref{Geff}) however arises as a purely classical
effect: it is weakly dependent on the distance scale ${\cal R}$ at
which the probe is deflected by the shock wave, because the probe
interacts with the fields of many KK gravitons in addition to the
ultralight mode. As distance increases, more intermediate
gravitons drop out from (\ref{fzeroone}) due to Yukawa suppression
from their mass, and $G_{eff}$ changes with distance, weakly and
in almost discrete jumps, thanks to the fact that the intermediate
mode functions change from a logarithm to a Yukawa-suppressed form
very sharply (\ref{asympqmass}). After every distance step of
$\sim 2 \ell/n$, $G_{eff}$ would jump by $\frac{\Delta
G_{eff}}{G_{eff}} \sim n \frac{L^2}{2\ell^2} \sim \frac{L^2}{\ell
{\cal R}}$, staying practically constant for a long distance
thereafter. The jumps are the largest at short scales, but
$\frac{\Delta G_{eff}}{G_{eff}} \la \frac{L}{\ell}$ as long as
${\cal R} > L$.

Note that if we view (\ref{Geff}) as a series in the powers of
$L/\ell$, and substitute the formula for $g_1^2$ (\ref{coupl})
also expanded in the powers of $L/\ell$, using
(\ref{modezerocoupls}) to extract the leading term we would obtain
the expansion
\be G_{eff} = \frac{1}{M_5^3 L} \, \Bigl(1 + \ldots\Bigr) \,
\Bigl( 1 + \frac12 \sum_{n=2}^{m_{max} < 2/{\cal R}}
\frac{L}{\ell} mL + \ldots \Bigr) \, , \label{Geffexp} \ee
where the factor $\Bigl(1 + \ldots\Bigr)$ out front stands for the
higher order $L/\ell$ corrections to $g_1^2$. Note that the
normalization of this expansion is by $M^2_4 = M^3_5 L$, precisely
the formula for the Planck mass in RS2 \cite{rs2}. We will discuss
it in more detail in the following sub-section.

From (\ref{Geff}) and the discussion leading to it we conclude
that in this regime $4D$ gravity is {\it composite}: the shock
wave behaves as a scattering amplitude generated by an exchange of
a resonance composed from all intermediate gravitons with masses
$m < 2/{\cal R}$, or {\it gravipartons}. The dominant contribution
comes from the ultralight mode. The $4D$ effective gravitational
coupling $G_{eff}$ runs with scale, changing in
well-separated sharp small jumps as long as $L \ll \ell$, because
the intermediate gravitons are very weakly coupled and very light.
Scattering of a probe particle against the shock probes the
internal structure of the graviton resonance. As the distance
increases towards $\ell$, new distance dependent corrections
induced by the $AdS_4$ curvature will become important, lifting
the leading order shock wave to its $AdS_4$ form and obscuring the
distance-dependent corrections generated by the intermediate
gravipartons.

We note that except for the coupling suppression of intermediate
and heavy modes,  manifest in (\ref{modemasscoups}), this regime
is analogous to the dynamics of gravity on a space with a large
compact dimension at distances between the fundamental Planck
length and the compactification radius \cite{add}. As one
increases the distance in the noncompact space, more and more KK
modes drop out due to their masses, and the gravitational
potential approaches the large distance limit set by the remaining
light mode.

\subsubsection{Ultralight Graviton Dominance}

At distances $\ell \la {\cal R} < {\ell}^3/L^2$ all the gravitons
except the ultralight one have decoupled. Their contribution to
(\ref{shockseriesg}) is exponentially small, due to Yukawa
suppressions from their masses, as is clear from Eq.
(\ref{asympqmass}). The shock wave at these distances is
completely composed of the ultralight mode, or in other words, the
$4D$ graviton resonance is extremely narrow. The effects of the
graviton mass are very small, as is clear from either of the Eqs.
(\ref{qone}) or (\ref{asympqzero}). The shock wave profile is very
closely approximated by
\begin{equation}
f_{\rm ultralight}(0, {\cal R} \ga \ell) = - \, \frac{p}{\pi} \,
g_1^2 \, \Big( 1 - {\cal O}(1) \frac{L^2}{\ell^2} \chi \Bigr) \,
Q_{1}\left(\cosh\chi\right) \, + \ldots \, . \label{leadshock}
\end{equation}
The leading $\chi$ dependence in (\ref{leadshock}) is { identical}
to the $AdS_4$ shock wave of Sfetsos (\ref{4dshock}), provided
that we identify $g_1^2$ with the $4D$ Planck mass $M_4$, $g_1^2 =
\frac{1}{M^2_4}$. In light of Eq. (\ref{coupl}) this yields
\be M^2_4 = \frac{2 M^3_5 \ell}{2\,\nu+1}
\frac{[-\,\partial_\nu\,P_\nu^{-1}\,\left(-\cos(y_0)\right)]}
{P_\nu^{-2}\left(-\cos(y_0)\right)} \Big|_{\nu= \nu_0} \, ,
\label{planckm} \ee
where $\nu_0$ is given by the ultralight mode mass according to
$\nu_0 = [\sqrt{9+4\ell^2\,m^2}-1]/2$. We note that this
definition is consistent with common practice in $GR$, where one
identifies the Planck scale $M_4^{-1}$ with the tree-level
coupling of the lightest mode in the spectrum. In $GR$ with a
massless graviton this definition coincides with the one where the
Planck scale is set by the gravitational scattering amplitude of
two particles in the limit when their separation goes to infinity.
Defined in this way, the Planck mass is a mere fixed dimensional
parameter obtained as a limiting value of $G_{eff}^{-1/2}$ in the
far infrared. This is still true in locally localized gravity:
once we have defined $M_4$ as in Eq. (\ref{planckm}), it is a
dimensional parameter which does not change any more. Indeed all
the coupling constants as defined in (\ref{coupl}) are independent
of momentum transfer or impact parameter. Any deviation of any
given mode from the $4D$ shock wave form comes only by courtesy of
its mass.

In the $L \ll \ell$ limit, the equation (\ref{planckm}) should be
understood as a perturbative result, obtained by the resummation
of all the contributions to the shock wave profile that retain the
{\it same} functional form in $\chi$, namely those $\propto
Q_1(\cosh \chi)$. These corrections communicate to the graviton
the presence of the IR cutoff set by the $AdS_4$ curvature radius
$1/\ell$: in locally localized gravity, the RS2 expression for
$M_4$ receives higher order IR corrections whose leading terms
found  by expanding (\ref{planckm}) are
\be M^2_4 = M^3_5 L \, \Bigl(1 + \frac{5}{12} (\frac{L}{\ell})^2 +
\ldots \Bigr) \, . \label{plrg} \ee
Further corrections can be easily  extracted from (\ref{planckm}).
Such terms may provide for useful
explicit checks of the dual
$CFT$ description.

Note that although the heavier gravitons have decoupled, the small
graviton mass still yields a very weak distance running of
$G_{eff}$, still defined -- following (\ref{Geff}) -- as the
coefficient of $Q_1$ in (\ref{leadshock}). However in this regime,
the discrete slow jumping that dominated the changing of $G_{eff}$
previously is replaced by a very slow continuous spatial
variation, gently weakening gravity according to $\Bigl(1 - {\cal
O}(1)\frac{L^2}{\ell^2} \frac{\cal R}{\ell}\Bigr)$. These
corrections remain tiny until ${\cal R}$ reaches $ {\ell}^3/L^2$.

\subsubsection{A Peek into $AdS_5$}

Once we go to distances ${\cal R} > \ell^3/L^2$, the corrections
from the ultralight graviton mass start altering the shock wave
more significantly. Indeed from the asymptotics of the ultralight
mode (\ref{asympqzero}) we see that the shock wave
(\ref{shockseries}) approaches
\begin{equation}\label{largechi}
f \left(0,\,\chi\gg 1\right) = \frac43 \frac{p}{\pi M^2_4} \,
e^{-\left(2+ {m^2 \ell^2}/{3} \right) \, {\cal R}/{\ell}} \, ,
\end{equation}
where we use $g^2_1 = 1/M^2_4$ of Eq. (\ref{planckm}). Comparing
this to the leading large distance behavior of the $4D$ $GR$ shock
wave on $AdS_4$ (\ref{4dshock}),
\begin{equation}\label{4dlargechi}
f_{4D\, AdS} \left(0,\,\chi\gg 1\right) = \frac43
\frac{p}{\pi\,M^2_4}\, e^{-2 \, {\cal R}/{\ell}} \,.
\end{equation}
we note the {\it extra} factor $\exp({-\frac{m^2 \ell^2}{3}
\,\frac{\cal R}{\ell}})$. It arises from the interplay of the
small mass and $AdS_4$ curvature, noted in Eqs. (\ref{qdeq}),
(\ref{qchilargem}). So if we compare it with the $4D$ form
(\ref{4dlargechi}), extracting $G_{eff} = f/f_{\rm standard}$ as
before, we find
\be G_{eff} \simeq  \frac{1}{M_2^4} \, e^{-({m^2 \ell^2}/{3})
\,{\cal R}/{\ell}} \, . \label{rfactor} \ee
Thus probing the shock with deflections of probes very far away
would suggest that the effective gravitational coupling continues
to run, with distance along the brane, eventually going to zero at
infinity. In the bulk this appears as a purely classical effect.

This scale dependence opens a small window into the fifth
dimension. To see it, we compare (\ref{largechi}) with the large
radius behavior of the solutions of a Klein-Gordon equation with a
small mass in $AdS_{d+1}$. That is sufficient because, due to
Lorentz contraction, the shock wave profile in $d+1$ dimensions
behaves like the gravitational potential of a source at rest in
$d$ dimensions. To find the scaling it is enough to consider large
distance limit of solutions of a massive Klein-Gordon in
$AdS_{d+1}$ Poincare patch, given by the metric $
ds_{{{AdS_{d+1}}}}^2= \frac{r^2}{\ell^2}(d\vec x^2_{d-1} - dt^2) +
\frac{\ell^2}{r^2} dr^2$, and ignore the $\vec x, t$ dependence,
so that the massive Klein-Gordon equation reduces to\footnote{For
exact expressions for scalar propagators in $AdS$, see
\cite{d'hoker} and references therein.} $
\left[\frac{r^2}{\ell^2}\,
\partial_r^2+\left(d+1\right)\,\frac{r}{\ell^2}\,
\partial_r-m^2\right]\,\Phi =0$.
Its regular solution at infinity scales as $r^{-d-m^2\ell^2/d}$
for $m\,\ell \ll 1$. In terms of the physical distance ${\cal R} =
\ell \log(r/\ell)$, this yields
\begin{equation}
\Phi_d \simeq e^{-\left(d+m^2\ell^2/d\right)\,{\cal R}/\ell }\,.
\end{equation}
The lesson is that both the leading and the subleading decay rates
of $\Phi$ are controlled by the {\it same} number $d$, counting
the spatial dimensions of $AdS$.

In contrast we see that in $AdS_4 \subset AdS_5$ framework, the
shock wave is a hybrid:
\begin{equation}
f\simeq e^{-\left(2+m^2\ell^2/3\right)\,{\cal R}/\ell}\, .
\end{equation}
The leading order decay rate is controlled by $AdS_4$ asymptotics,
and is the same as for conventional massless $GR$ in $AdS_4$. The
subleading order probes the full $AdS_5$ geometry, making long
range gravitational force {\it weaker} than in $4D$ $GR$ on
$AdS_4$. However it does so a bit more {\it slowly} than a
perturbative graviton mass term would have done in $AdS_4$ alone,
where as we have seen above the subleading correction to the decay
rate would have been $\propto \exp[-(m^2 \ell^2/2){\cal R}/\ell]$.
In other words, the rate of $G_{eff}$ running discerns the
influence of asymptotic $AdS_5$. Note that this scaling is very
similar with the scaling of the operator which acquires anomalous
dimension in the dual $CFT$, which is responsible for the
emergence of the graviton mass \cite{defect}.

\subsection{Recovering Minkowski} \label{road}

As we have already seen from (\ref{aiseshock}), the shock wave
(\ref{master}) reduces in the leading order to the Aichelburg-Sexl
solution at distances $L \ll {\cal R} \ll \ell$. The limit to RS2
corresponds to fixing $L$ and taking $\ell \rightarrow \infty$. We
also need to rescale $\chi = {\cal R}/\ell$ and keep ${\cal R}$
fixed while sending $\ell$ to infinity. In this case the
background (\ref{metric}) reduces to RS2, and so the Ansatz for
the shock wave metric (\ref{wave}) goes to the form obtained by
boosting a mass confined to the RS2 brane \cite{roberto}. Our
shock wave profile also reduces precisely to Emparan's solution as
well. Of course, this is expected because of the symmetries of the
theory and the limiting form of the masses and couplings (see
Appendix C), which ensure that the spectrum reduces back to the
RS2 one. Here we outline this limit, leaving more details for
Appendix D.

We start with (\ref{shockseriesg}) and separate the ultralight
mode $\propto g_1^2$ from the rest. When $\ell \rightarrow
\infty$, the coupling reduces to the RS2 value $\frac{1}{M_5^3 L}
= \frac{1}{M_4^2}$ and the mode function $Q_{\nu_1}(\chi)$ becomes
exactly $Q_{\nu_1} \left(\cosh\chi\right) \rightarrow -\log({\cal
R} /2\ell)-1$ because the higher order terms in (\ref{apprq}) are
proportional to ${\cal O}(({\cal R}/\ell)^2)$ and so they vanish
when $\ell \rightarrow \infty$, while $\psi(2) + \gamma_E = 1$.
Further, we can replace the normalization factor $1/\ell$ in the
logarithm by $1/L$ by using a $4D$ diffeomorphism along the brane
\cite{thooft}. Thus the ultralight mode contribution yields
\begin{equation}
f_{A-S} =\frac{p}{\pi\,M_4^2}\,\log\frac{\cal R}{2L}\, + \,
\frac{p}{\pi\,M_4^2}. \label{zeroas}
\end{equation}
The sum of the massive modes reduces to an integral, thanks to
properties of hypergeometric functions (see Appendix D.),
\begin{equation}
- \frac{p}{\pi} \lim_{\ell \rightarrow \infty, {\cal R} < \ell}
\sum_{n=2}^{\infty} g^2_n Q_{\nu_{n}}(\chi) = -\frac{p}{2\pi
M_5^3\,L}\,\int_0^{\infty} \,\frac{dm}{m}\,\frac{4}{\pi^2}\,
\frac{K_0\left(m {\cal
R}\right)}{J_1\left(mL\right)^2+Y_1\left(mL\right)^2} \,.
\label{robertos}\end{equation}
Adding (\ref{zeroas}) and (\ref{robertos}) we obtain the RS2 shock
wave on the brane found by Emparan \cite{roberto} (where we use
slightly different normalizations):
\be f_{RS2} = \frac{p}{\pi\,M_4^2}\,\log\frac{\cal R}{2L}\, + \,
\frac{p}{\pi\,M_4^2} \, - \, \frac{p}{2\pi
M_5^3\,L}\,\int_0^{\infty} \,\frac{dm}{m}\,\frac{4}{\pi^2}\,
\frac{K_0\left(m{\cal
R}\right)}{J_1\left(mL\right)^2+Y_1\left(mL\right)^2} \,.
\label{rs2shock} \ee
This is another nice check of the calculation.

It is instructive to scrutinize this limit more closely. Expanding
the integrand in powers of $mL$ we get
\begin{equation}
f_{RS2} = \frac{p}{\pi\,M_4^2}\,\log\frac{\cal R}{2L}\, + \,
\frac{p}{\pi\,M_4^2} \,  - \, \frac{p}{2\pi M^2_4}
\frac{L^2}{{\cal R}^2} \,  +  \, \ldots \, , \label{rs2expansion}
\end{equation}
where we only keep the leading correction ${\cal O}(L^2/{\cal
R}^2)$. This is the surviving correction from the intermediate
gravitons in the limit $\ell \rightarrow \infty$. To see this,
observe that in the limit $\ell \rightarrow \infty$ we can replace
the sums in (\ref{fzeroone}) with
\be \sum_{n = 2}^{m<2/{\cal R}} \rightarrow
\int_{1/\ell}^{2/\cal{R}} \, \ell dm   \, . \ee
We regulate all mass integrals in the IR by the cutoff $\mu
=1/\ell$ since we only work in the box ${\cal R} < \ell$. This
properly encodes the asymptotic behavior of the intermediate mass
modes $Q_{\nu_n}$. Using (\ref{modemasscoups}) as $\ell
\rightarrow \infty$, the correction terms in (\ref{fzeroone})
become
\begin{eqnarray}
{\cal C} &=& \frac{p}{\pi} \,  \Bigl(  g_1^2 \, [\psi(\nu_1+1) +
\gamma_E] + \sum_{n=2}^{m_{max} <2/{\cal R}} g_n^2 \,
[\psi(\nu_n+1) +
\gamma_E] \Bigr) = \nonumber \\
&& ~~~~~~~~~~~~~~~~~~~~~~~~ = \frac{p}{\pi M_4^2} \Bigl( 1 +
\frac12 \int_{1/\ell}^{2/{\cal R}} L^2\,m\,dm \ln(\ell m) \Bigr) \, ,
\label{corrlimit}
\end{eqnarray}
where we have used $\psi(\nu_n+1) + \gamma_E = \sum_{k=1}^n 1/k =
\int^m_{1/\ell} \frac{dm}{m} = \ln(\ell m)$. Direct integration
yields
\be {\cal C} =  \frac{p}{\pi M_4^2} \Bigl( 1 - \frac{L^2}{{\cal
R}^2} \ln(\frac{\cal R}{2\ell}) - \frac12 \frac{L^2}{{\cal R}^2} +
\, \ldots \Bigr) \, , \label{lim1} \ee
where we omit all the terms which vanish in the limit $\ell
\rightarrow \infty$. We note that the second term has logarithmic
divergences in the IR. On the other hand, we also find that Eq.
(\ref{Geff}) in this limit reduces to
\be G_{eff} = g_1^2 \,  \Bigl( 1 + \frac12 \int_{1/\ell}^{2/{\cal
R}} dm\, m L^2 + \ldots \Bigr) = \frac{1}{M_4^2}  \, \Bigl( 1 +
\frac{ L^2}{{\cal R}^2} + \ldots \Bigr)  \, , \label{Geffint} \ee
again ignoring all the terms $\propto L/\ell$ that vanish as $\ell
\rightarrow \infty$. Plugging (\ref{lim1}), (\ref{Geffint}) back
into (\ref{fzeroone}), we see that the $G_{eff} \ln({\cal
R}/{2\ell})$ contains another logarithmic divergence $ \propto
\frac{L^2}{{\cal R}^2} \ln(\frac{\cal R}{2\ell})$, but of opposite
sign. Thus the logarithmic divergences cancel exactly, including
the finite logarithmic pieces, and the surviving form of the shock
is precisely the long distance limit of the RS2 shock
(\ref{rs2expansion}), including the correct coefficients of all
the leading terms.

This confirms our interpretation of slow jumpy changes in $G_{eff}
$ as a running induced by the $CFT$ loops in the dual theory.
Indeed we see that in the RS2 limit, the running precisely cancels
the IR divergence coming from the IR contributions of the lightest
$CFT$ excitations, and the surviving subleading term in the shock
wave solution at long distances is the finite $CFT$ correction.

\section{Conclusions}

The bulk picture painted with our shock waves gives us glimpses of
the dual $AdS/CFT$ interpretation of locally localized gravity in
terms of a defect $CFT$ \cite{kara2,defect}. Imagine that a fat
defect, of size $L$ and codimension one, has been built out of
$4D$ $CFT$ degrees of freedom, as proposed in \cite{kara2,defect}.
On the defect, the $CFT$ is cut off in the UV at $1/L$, set by the
thickness of the defect. If the defect is stable, or at least very
long lived, its low energy excitations will be non-destructive
because they correspond to a change of the internal structure of
the defect. They are described by an effective field theory
emerging from the dual $CFT$ mixing with the defect \cite{defect}.
The conformal symmetry of these excitations is broken down to a
$SO(3,2)$ subgroup at a low scale $1/\ell$, generated by the
strong coupling dynamics. This scale at the moment is not
calculable in a precise way, however there are compelling
arguments pointing out how it can emerge in several different ways
\cite{defect}. The discrete tower of intermediate bulk gravitons
with masses $m \sim n/\ell < 1/L$ corresponds to the defect
excitations, and their dynamics has been described by an effective
$CFT$ in $AdS_4$ \cite{porrati}. The heavy gravitons with masses
above $1/L$ are not a part of the $CFT$ dual, but only emerge
after we specify a UV completion of the cutoff, and in the case we
consider they follow from the assumption that the bulk picture can
be used all the way up to the $5D$ Planck scale $M_5$.

After we impose the UV cutoff $1/L$, the local geometry around the
defect becomes dynamical, described by $4D$ gravity whose Planck
mass arises from the finite UV cutoff. Once the defect excitations
and the IR cutoff $\ell$ are included, there will be additional
terms in the vacuum polarization amplitude of the graviton, as
argued in \cite{kara2,porrati,defect}, which will also receive
contributions which depend on the external momentum, on the masses
of the $CFT$ excitations and on the IR cutoff $1/\ell$. In the
effective $CFT$ on $AdS_4$ picture, these latter terms will come,
for example, from the lower limit of momentum integration over the
defect excitations in the loops. In this description of the defect
dynamics, the new terms influence the graviton dynamics in several
different ways. The corrections from the IR cutoff, by conformal
symmetry, will yield a correction to the graviton 2-point
function, scaling as $1/\ell^4 $, which is weighed by the two
powers of the graviton coupling $1/M_4$ and multiplied by the
number of $CFT$ degrees of freedom. This gives $m^2 \sim g_*
\frac{1}{M^2_4} \frac{1}{\ell^4} \sim \frac{L^2}{\ell^4}$, exactly
the ultralight graviton mass \cite{porrati}. In the defect $CFT$
this is interpreted as the statement that some operator acquires a
small anomalous dimension in the presence of the defect
\cite{defect}. Further these terms will also correct the formula
for the $4D$ Planck scale, defined as the coupling of the
ultralight mode, which should really be $M^2_4 \sim g_*/L^2 (1 +
O(L^2/\ell^2))$, where $g_* = M^3_5 L^3$ is the number of $CFT$
degrees of freedom and $O(L^2/\ell^2)$ are the higher order
quantum corrections from the defect excitations deformed at the IR
cutoff $\ell$. Indeed from our exact shock wave calculation
(\ref{planckmass}) we find that $M_4^2 = M^3_5 L \Bigl(1
+\frac{5}{12} (\frac{L}{\ell})^2 + \ldots \Bigr)$, in agreement
with the estimate based on effective field theory.

The corrections which depend on the external momentum and masses
of defect excitations are again in agreement with the description
of the defect by an effective $CFT$ in $AdS_4$ \cite{porrati}. The
defect excitations in the graviton vacuum polarization diagrams
renormalize the Newton's constant $G_{eff}$, dressing up the
coupling and making the graviton look composite at distances below
$\ell$. We have seen how this works explicitly by taking the limit
of $\ell \rightarrow \infty$, where we restore the RS2 shock wave
found by Emparan \cite{roberto}, where the excitation-induced
running of $G_{eff}$ precisely cancels the logarithmic
singularities in the subleading long range contributions from the
intermediate modes, and renders the shock finite. The surviving
terms are precisely the remaining $CFT$ vacuum polarization
corrections to the graviton at large distances.

Because of the details of the embedding of the $AdS_4$ brane in
the bulk, from the bulk point of view our relativistic projectile
moves along a trajectory which is {\it not} a null geodesic in the
full bulk. Since the particle is stuck on the brane, there is a
force induced by the brane's structure which acts on it, and so
its trajectory is deformed away from a bulk null geodesic. From
the dual $CFT$ point of view, a shock wave of a null particle in
the bulk is interpreted as an excitation of the light cone states
of the dual $CFT$ \cite{horit}. Hence the shock wave of a source
on the brane should correspond to the light cone states
``fattened" by the interactions with the defect. Far from the
source the defect excitations decouple because of the gap in their
spectrum, and the only signature of the defect is the small mass
of the ultralight graviton, which drives the shock field to zero
faster than in $4D$ $GR$ (but more slowly than a perturbative
graviton mass term would have done in $AdS_4$). If we extract the
effective Newton's constant far from the source by comparison with
the solution in $AdS_4$, we see that it goes to zero as a function
of distance at a rate set by the anomalous dimension of the
operator which should be dual to the graviton in $AdS_4 \subset
AdS_5$. This points to a connection with the holographic rescaling
in the boundary $CFT$ \cite{lenny}, and it would be interesting to
explore further. Note that the most dramatic deviations from the
$4D$ become noticeable through the subleading corrections to the
shock wave profile only at distances ${\cal R} > \ell^3/L^2$ along
the brane.

In sum, in this paper we have presented the exact solution for the
gravitational field of a relativistic particle living on an
$AdS_4$ brane in $AdS_5$ space. The solution is the first exact
background in the framework of locally localized gravity in $AdS$
braneworlds. As advocated in \cite{kallet} the shock wave is a
powerful tool for investigating the nature of the transverse
traceless graviton sector. From exploring the solution we find the
precise formula for the $4D$ Planck mass relating it to the $5D$
Planck mass and the curvature radii. It includes higher order
corrections that depend on the $AdS_4$ radius $\ell$, and reduces
to the RS2 expression as $\ell \rightarrow \infty$. In the dual
$CFT$ description, this is interpreted as corrections from the IR,
where strong coupling effects in the defect $CFT$ manufacture the
IR scale $1/\ell$. This is also consistent with holographic
arguments suggested in \cite{bura}, who noted that the dual
description of the $AdS_4 \subset AdS_5$ system can only cover a
portion of the bulk $AdS_5$ space transverse to the $AdS_4$ brane.
Our shock gives an interesting explicit handle for exploring these
IR modifications in the dual $CFT$. We also confirm that the shock
wave looks $5D$ at short distances and approximates very closely
the $4D$ $AdS$ shock wave in the leading order. However at
distances ${\cal R} \ga \ell^3/L^2$ along the brane, the
subleading terms become important because they induce an extra
correction to the decay rate of the graviton wave function along
the brane. The correction knows of the presence of the extra
dimension, and leads to subleading deviations from the $AdS_4$
shock wave of \cite{kostas}, which a $4D$ observer may interpret
as a spatial ``running" of the $4D$ effective gravitational
coupling. Following this variation of $G_{eff}$ at large ${\cal
R}$ one would be able to count up all spatial dimensions without
ever moving away from the brane. In \cite{bura} the authors offer
a qualitative argument based on light-sheet holography, suggesting
that gravity may remain $4D$ all the way out to infinity if it is
probed over times shorter than $\ell$. We find a somewhat stronger
version of this statement, which is that gravity deviates away
from $4D$ $GR$ in a significant way only at distances $\ga
\ell^3/L^2$ along the brane.

In our solution the scalar graviton mode, present whenever there
is a mass term \cite{higuchi}, is conspicuously missing. The
reason is the same as the one encountered in the construction of
shock waves in DGP braneworlds \cite{kallet}. Relativistic
perturbations on the brane have a vanishing stress-energy tensor
trace, $T^{\mu}{}_\mu = 0$, and therefore do not source the scalar
graviton field. This logic is consistent barring the problems with
strong coupling, which may be under control here thanks to the
perturbation theory results of \cite{Kogan,kkr,nimags}. One may be
able to test this explicitly by considering very fast particles
with non-zero rest mass, moving with a speed ${\tt v} \rightarrow
c$ as sources, and including the effect of their mass. They would
source the scalar graviton, which should be a perturbation on top
of the shock wave, suppressed by the source mass-to-momentum ratio
$M/p = \sqrt{1/{\tt v}^2-1} \ll 1$ \cite{kallet}.
It would be interesting to investigate the perturbative dynamics
of the scalar and its implications for locally localized gravity
and for its dual description in more detail.

\acknowledgments We thank Thibault Damour, Roberto Emparan, Ted
Jacobson, Andreas Karch, Konstadinos Sfetsos and John Terning for
useful discussions. N.K. is grateful to IHES, Bures-sur-Yvette,
France, for kind hospitality during this work. L.S. acknowledges
the kind hospitality of the Theory Division, CERN, during the
course of this work. This work was supported in part by the DOE
Grant DE-FG03-91ER40674, in part by the NSF Grant PHY-0332258 and
in part by a Research Innovation Award from the Research
Corporation.

\appendix

\section{Derivation of the Wave Profile Field Equation}

To evaluate $G^A_{5\,B}$ in (\ref{fieldeqs}) we can use the same
conformal transformation trick as in \cite{kallet}, splitting the
metric as $ds^2 = \Omega^2 d\bar s^2$, where
\begin{equation}
d\bar s^2 = \frac{4\,du\,dv}{ (1-uv/\ell^2 )^2} - \frac{4
\delta(u) f du^2}{(1- uv/\ell^2)^2} +\ell^2 \,
\Bigl(\frac{1+uv/\ell^2}{1-uv/\ell^2} \Bigr)^2 \,
(d\chi^2+\sinh^2\chi\,d\phi^2 )+dz^2 \, . \label{confwave}
\end{equation}
We could now proceed with a straightforward albeit tedious
computation of the Einstein tensor for $d\bar s^2$, treating
$\delta(u)$ and its derivatives as distributions. We can shorten
the labor, however, by noting that the metric (\ref{confwave}) is
related to the conformal de Sitter metric, Eq. (12) of the second
of references  \cite{kallet}, by a substitution
\be \frac{1}{\ell} = i H \, , \qquad \qquad \chi = i \theta \, .
\label{wick} \ee
Hence we can take the components of the Ricci tensor computed
there and map them back, finding
\begin{eqnarray}
\bar R_{5\,uu} &=& 2 \delta(u) \Bigl(\partial_z^2 f +
\frac{1}{\ell^2} ( \Delta_2 f + 4f) \Bigr) \, , \nonumber \\
\bar R_{5\,uv} &=& -\frac{6}{\ell^2 (1- uv/\ell^2)^2} \, ,
\nonumber \\
\bar R_{5\,ab} &=& - 3 \, \Bigl(\frac{1+uv/\ell^2}{1-uv/\ell^2}
\Bigr)^2 g_{ab} \, , \label{ricciconf}
\end{eqnarray}
where $\bar R_{5\,AB}$ is the Ricci tensors of (\ref{confwave})
and $a,b$ and $g_{ab}$ are the indices and the metric on the
``unit" ${\cal H}_2$, respectively. The operator $\Delta_2$ is the
Laplacian on the ``unit" ${\cal H}_2$, where we have properly
accounted for the sign flips from the map (\ref{wick}). Now we
``invert" the conformal transformation $ds_5^2 = \Omega^2 d\bar
s_5^2$ to find the Ricci tensor for $ds^2$, given by
\begin{equation}
R_{5\,AB} = \bar R_{5\,AB} - 3 \bar \nabla_A \bar \nabla_B \ln
\Omega + 3 \bar \nabla_A \ln \Omega \bar \nabla_B \Omega  - \bar
g_{AB} \Bigl(\bar \nabla^2 \ln \Omega + 3 (\bar \nabla \ln
\Omega)^2 \Bigr) \, , \label{conftrans}
\end{equation}
which, with $\Omega(|z|) = \Bigl[{L}/[\ell
\sin(\frac{|z|+z_0}{\ell})]\Bigr]$, and using distributional rules
for the derivatives of $\delta$-functions, e.g. $u \delta(u) = 0$,
$u^2 \delta^2(u) = 0$ and $f(u) \delta'(u) = - f'(u) \delta(u)$
\cite{thooft,kostas}, yields
\begin{eqnarray}
R_{5\,zz} &=& -4 \, \partial_z \Bigl(\frac{\partial_z
\Omega}{\Omega} \Bigr)  \, ,  \nonumber \\
R_{5\,uu} &=& 2 \delta(u) \Bigl(\partial_z^2 f +
3\frac{\partial_{|z|} \Omega}{\Omega} \partial_{|z|} f +
\frac{1}{\ell^2} (\Delta_2 f - 2f) \Bigr) + 4 \partial_z
\Bigl(\frac{\partial_z
\Omega}{\Omega} \Bigr) \, f \delta(u)  \, , \nonumber \\
R_{5\,uv} &=& -\frac{2}{(1-uv/\ell^2)^2} \,
\partial_z \Bigl(\frac{\partial_z \Omega}{\Omega} \Bigr) \, , \nonumber \\
R_{5\,ab} &=& \ell^2 \Bigl(\frac{1+uv/\ell^2}{1-uv/\ell^2}
\Bigr)^2    g_{ab} \, \partial_z \Bigl(\frac{\partial_z
\Omega}{\Omega} \Bigr)  \, . \label{riccicomp}
\end{eqnarray}
From the definition of $\Omega$, we have
\begin{eqnarray}
\frac{\partial_z \Omega}{\Omega} &=& - \frac{2\Theta(z)-1}{\ell}
\, \cot[(|z|+z_0)/\ell\,] \, , \nonumber \\
\partial_z \Bigl(\frac{\partial_z \Omega}{\Omega}\Bigr) &=& - 2
\frac{\cot[(|z|+z_0)/\ell]}{\ell} \delta(z) + \frac{1}{{\ell}^2
\sin^2[(|z|+z_0)/\ell\,]} \, . \label{omegaterms} \end{eqnarray}
Clearly, in the presence of the shock wave the bulk is not $AdS_5$
any more, since the waves extend off the brane and deform the
bulk. This can be seen from the last term of $R_{5\,uu}$, and also
from the Riemann tensor. Now, we can rewrite (\ref{fieldeqs}) as
\begin{eqnarray}
R^{A}_{5\,B} &=& -\frac{4}{L^2} \, \delta^A{}_B  + 2 \,
\Bigl(\frac{1}{L^2} \, - \frac{1}{\ell^2}\Bigr)^{1/2} \, \delta(z)
\Bigl( 4 \delta^A{}_B -3 \delta^\mu{}_\nu \, \delta^A_\mu \,
\delta^\nu_B \Bigr) \nonumber \\
&& ~~~~~~~~~~~~ + 2 \, \frac{p}{M_5^3 \ell^2} \,
\frac{1}{\sinh\chi} \, \delta^\mu_v \, \delta^u_\nu \,
\delta^A_\mu \, \delta^\nu_B \, \delta(\chi) \, \delta(\phi) \,
\delta(u) \, \delta(z) \, . \label{riccieqs}
\end{eqnarray}
The new equation for the wave profile $f$ comes from the ${uu}$
component of (\ref{riccieqs}). Using $\delta(\chi)/\sinh\chi =
\delta(\cosh \chi - 1)$, after a straightforward calculation we
find that the {linear} field equation for the wave profile is
\begin{equation}
\partial^2_z f +
3\frac{\partial_{|z|}\Omega}{\Omega}\,\partial_{|z|}f+
\frac{1}{\ell^2} \,( \Delta_2 f -2 f) = \frac{2p}{M^3_5 \ell^2}
\,\delta\left(\cosh\chi-1\right)
\,\delta\left(\phi\right)\,\delta\left(z\right) \label{feqnap}
\end{equation}
The remaining components of (\ref{riccieqs}) are trivially
satisfied by virtue of the background equation $L/\ell =
\sin(z_0/\ell)$.

We note that in the case of $AdS$ backgrounds the procedures to
solve this equation are more straightforward (albeit technically
more involved) that in de Sitter background. The reason is that
the space transverse to the source of the shock wave is
non-compact, and so implementing the boundary conditions for the
single particle problem is simpler, as there do not arise
additional singularities that appear on a 2-sphere in de Sitter
\cite{hota,pogri,kostas}.

\section{Series Representation of the Shock Wave}

Here we derive the series representation of the shock wave, Eq.
(\ref{shockseries}) from the integral formula (\ref{master}).
Using Eq. (3.3.1 (8)) of \cite{bateman} we write
\begin{equation}
P_\nu\left(z\right)=\frac{\sin\left(\pi\,\nu\right)}{\pi\,\cos\left(\pi\,\nu\right)}
\,\left[Q_\nu\left(z\right)-Q_{-\nu-1}\left(z\right)\right]\, .
\label{pqq}
\end{equation}
After doubling up the domain of integration in (\ref{shockseries})
according to $\int_0^\infty=(1/2)\int_{-\infty}^\infty$, using
(\ref{pqq}) we rewrite it as
\begin{eqnarray}
f\left(0,\,\chi\right) &=& \frac{p}{2\pi M^3_5
\ell}\frac{1}{2\,\pi\,i}\int_{-\infty}^\infty dq\,q
\frac{P_{iq-1/2}^{-2}\left(-\cos(y_0)\right)}
{P_{iq-1/2}^{-1}\left(-\cos(y_0)\right)} \times \nonumber \\
&& ~~~~~~~~~~~~~~~~~~~~~~ \times
\left[Q_{iq-1/2}\left(\cosh\chi\right)-Q_{-iq-1/2}\left(\cosh\chi\right)\right].
\end{eqnarray}
To evaluate the integral, we split it into two parts setting $f =
f^+ + f^-$, where
\begin{eqnarray}
f^+ &=& \frac{p}{2\pi M^3_5
\ell}\frac{1}{2\,\pi\,i}\int_{-\infty}^\infty dq\,q
\frac{P_{iq-1/2}^{-2}\left(-\cos(y_0)\right)}
{P_{iq-1/2}^{-1}\left(-\cos(y_0)\right)}
Q_{iq-1/2}\left(\cosh\chi\right) \, , \nonumber \\
f^- &=& \frac{p}{2\pi M^3_5
\ell}\frac{i}{2\,\pi}\int_{-\infty}^\infty dq\,q
\frac{P_{iq-1/2}^{-2}\left(-\cos(y_0)\right)}
{P_{iq-1/2}^{-1}\left(-\cos(y_0)\right)}
Q_{-iq-1/2}\left(\cosh\chi\right) \, .
\end{eqnarray}
Notice that by the reality of the Legendre functions, the two
integrals $f^{\pm}$ are complex conjugates of each other, if they
are convergent: $f^- = (f^+)^*$. Thus to evaluate $f$, it is
sufficient to only compute one of $f^{\pm}$. Let us compute
$f^{+}$. For large $q$, the integrand of this integral goes as
$\exp(-i\,q\,\chi)$ (see \cite{gradrhyz}, Eq. (8.723.2)).
Therefore we can close the integration contour on the half
$q$-plane where $Im[q]<0$. Defining $\nu \equiv iq-1/2$, the
integration yields
\begin{equation}
f^{+} =\frac{p}{4\,\pi\,M^3_5\,\ell}
\,\sum_{\nu>-1/2}\left(2\,\nu+1\right)\,
\frac{P_\nu^{-2}\left(-\cos(y_0)\right)}{\partial_\nu\,P_\nu^{-1}\,\left(-\cos(y_0)\right)}
\,Q_\nu\left(\cosh\chi\right)\,.
\end{equation}
where the sum runs over all the poles of the integrand, located at
the values of $\nu > -1/2$ for which
$P_\nu^{-1}\,\left(-\cos(y_0)\right)=0$. Because $f^+$ is finite
(except for the short distance singularities, of course) and real,
$f^- = f^+$, and hence $f = 2 f^+$, or
\begin{equation}
f\left(0,\,\chi\right) =\frac{p}{2\,\pi\,M^3_5\,\ell}
\,\sum_{\nu>-1/2}\left(2\,\nu+1\right)\,
\frac{P_\nu^{-2}\left(-\cos(y_0)\right)}{\partial_\nu\,P_\nu^{-1}\,\left(-\cos(y_0)\right)}
\,Q_\nu\left(\cosh\chi\right) \, , \qquad \nu = iq-1/2 \, .
\label{seriesap}
\end{equation}

\section{Spectrum of $AdS_4 \subset AdS_5$: Masses and Couplings}

Here we determine the masses and couplings of graviton modes in
the limit $L \ll \ell$. The modes are determined by the secular
equation (\ref{eignens}). A suitable expression for the Legendre
function $P_\nu^{-1}(x)$ is
\begin{equation}
P_\nu^{-1}\left(x\right)=\left(\frac{1-x}{1+x}\right)^{1/2}
\,F\left(-\nu,\,1+\nu;\,2;\,1/2-x/2\right)\, ,
\end{equation}
which means that the problem of finding the eigenmodes reduces to
finding the roots of the hypergeometric function. In the small
$y_0 = L/\ell + \ldots$ limit, we substitute $x=-\cos(y_0)\simeq
-1+y_0^2/2$ in the hypergeometric function  and expand it. The
series representation is
\begin{equation}
\frac{1}{\Gamma(2+\nu)\,\Gamma(1-\nu)}\,\left\{1+\nu(\nu+1)\,\frac{y_0^2}{4}
\,\sum_{j=0}^\infty\frac{(2+\nu)_j\,(1-\nu)_j}{(j+1)!\,j!}
\,\left[h_j-\log\frac{y_0^2}{4}\right]\left(\frac{y_0^2}{4}\right)^j\right\}
\, , \label{hyperg}
\end{equation}
where
\begin{equation}
h_j=\psi(1+j)+\psi(2+j)-\psi(2+\nu+j)-\psi(1-\nu+j) \, ,
\end{equation}
and $\psi(z)$ is the logarithmic derivative of the
$\Gamma$-function. We only consider the leading order expressions
for the masses, and so in the small $y_0$ limit, we can truncate
the series (\ref{hyperg}) to the leading order in $y_0$, which
means drop all the terms beyond $j=0$. In this limit the secular
equation reduces to
\begin{equation}
\nu(\nu+1)\, [\psi(2+\nu)+\psi(1-\nu) -\psi(1)-\psi(2)] =
\frac{4}{y_0^2} + \nu(\nu+1)\, \log\frac{y_0^2}{4} \,.
\label{dilogs}
\end{equation}
Further, in this limit we can ignore the logarithm on the RHS for
intermediate graviton modes $|\nu| < 1/y_0$. By analyticity of the
integration that produced the series representation of the shock
wave (\ref{seriesap}), we must require $\nu > -1/2$. Then the
large contribution from $y_0$, which grows as a second order pole,
can be compensated by the singularities of the $\psi$ function,
which has poles at negative integers and zero. Given the
constraint $\nu > -1/2$, we see that the only relevant term on the
LHS is $\psi(1-\nu)$. So whenever $\nu$ approaches a positive
integer, $\psi$ shoots up sufficiently quickly to match the
$1/y_0^2$ term. Setting $\nu=n+\varepsilon$ with $n=1,2,3, \ldots
$ and a small $\varepsilon$, we find $\psi(1-n-\varepsilon) =
1/\varepsilon + {\rm finite~terms}$. Substituting this in
(\ref{dilogs}), we can solve for $\varepsilon$, or equivalently
for $\nu_n$, to the leading order in $y_0 = L/\ell + \ldots$. We
find
\begin{equation}\label{nukpiccolo}
\nu_n = n+\frac{y_0^2}{4}\,n(n+1)\, \, + \ldots \,,\qquad n=1,2,3,
\ldots  \,.
\end{equation}
The masses of the spectrum, by $m^2 =
\frac{\nu\left(\nu+1\right)-2}{\ell^2}$, are therefore given by
\be m^2 = \frac{1}{\ell^2} \Bigl[(n-1)(n+2) + \frac{1}{2}
n(n+1/2)(n+1) \frac{L^2}{\ell^2} \Bigr] \, + \ldots \, , \qquad  n
= 1, 2, 3, \ldots \, .\label{eigenmassC} \ee
When $\varepsilon\sim n^2\,y_0^2/4$ ceases to be small, for $n\sim
2/y_0$, the above formula for the KK masses should be replaced.
This happens when the mass reaches $ m \sim 1/L$. At such high
scales, effects of the curvature can be ignored, and the system
behaves effectively as a box of size $2(\pi-y_0) \ell$. Starting
with the expansion valid for large $\nu$
\begin{equation}\label{asymp}
P_\nu^\mu\left(-\cos y_0\right)\simeq
\frac{\Gamma\left(\nu+\mu+1\right)}{\Gamma\left(\nu+3/2\right)}\,
\left(\frac{2}{\pi\,\sin y_0}\right)^{1/2}\,
\cos\left[\left(\nu+\frac{1}{2}\right)\left(\pi-y_0\right)+
\frac{2\mu-1}{4}\,\pi\right]\, ,
\end{equation}
in the secular equation (\ref{eigenmassC}) we find that the
eigenvalues in this regime are
\begin{equation}\label{largeeig}
\nu_n=-\frac{1}{2}+\frac{\pi/4+n\,\pi}{\pi-y_0} \, + \ldots \, =
n\left(1+y_0/\pi\right)\, + \ldots \,,\qquad n\ga 2/y_0\,.
\end{equation}
Therefore the heavy graviton masses are
\be m^2 = \frac{1}{\ell^2} \Bigl[ n^2 + n + \frac{2L^2}{\pi
\ell^2} n^2 \Bigr] \, + \ldots \, , \qquad n \ga \frac{\ell}{L}
.\label{eigenmasshC} \ee
Note that these modes are above the $CFT$ cutoff $1/L$, and hence
are not in the calculable regime of a dual $CFT$, but are still
well defined on the bulk side as long as $L \ll 1/M_5$ (and de
facto negligible at all but the shortest distances ${\cal R} <
L$).

We now determine the couplings of these modes. From the series
expression for the shock wave profile (\ref{shockseries}) we can
simply read off the couplings as the coefficients of
$Q_\nu\left(\cosh\chi\right)$:
\be g^2_n = - \frac{2\,\nu+1}{2 M^3_5 \ell}
\frac{P_\nu^{-2}\left(-\cos(y_0)\right)}{\partial_\nu\,P_\nu^{-1}\,\left(-\cos(y_0)\right)}
\Big|_{\nu= \nu_n} \, . \label{couplC} \ee
This formula is an exact bulk expression for couplings of each
individual graviton mode to the matter on the brane. The (-) sign
compensates the fact that
$\partial_\nu\,P_\nu^{-1}\,\left(-\cos(y_0)\right)$ is negative.

We next express the couplings as functions of the mode masses to
the leading order in $y_0 \ll 1$. We determine
$P_\nu^{-2}\left(-\cos(y_0)\right)$ from (\ref{legexpansion}) and
$\partial_\nu\,P_\nu^{-1}\,\left(-\cos(y_0)\right)$ using the
leading-order truncation of (\ref{hyperg}). After a
straightforward calculation this yields
\be g^2_n = \frac{L}{4 M^3_5 \ell^2} \, \frac{(2n+1)n(n+1)}{n+2}
\, \frac1{n-1+n(n+1)(L^2/4\ell^2)} \, . \label{modecouplings} \ee
Note that the coupling of the ultralight mode, $g_1 =
\frac{1}{M^2_4}$ is {\it much} stronger than that of the other
modes: it is
\be g^2_1 = \frac{1}{M^3_5 L}  \, + \ldots \, ,
\label{modezerocouplings} \ee
i.e. just the standard RS2 Newton's constant, whereas the
couplings of heavier modes are suppressed relative to it by powers
of $L/\ell$. In fact using (\ref{eigenmassC}) and omitting higher
order terms we can rewrite the couplings of heavy modes as
\be g^2_m \simeq  \frac{m L}{2 M^3_5 \ell} = \frac{1}{2M^2_4}
\frac{L}{\ell}{mL}\, . \label{modemasscouplings} \ee
As mass increases, this formula eventually saturates to $g^2_m
\simeq \frac{1}{M^3_5 \ell}$. This is reminiscent of the
tunnelling suppression of bulk modes in RS2 \cite{rs2,tunneling}.

\section{The Limit of RS2 Shock Wave}

Here we outline how to take the RS2 limit $\ell\rightarrow \infty$
and recover the result of the shock wave calculation of
\cite{roberto}. Our starting point is the series
representation~(\ref{seriesap}) of the shock wave profile
\begin{eqnarray}
&&f\left(0,\,\chi\right) =\frac{p}{2\,\pi\,M^3_5\,\ell}
\sum_{\nu>-1/2}\left(2\,\nu+1\right) \frac{P_\nu^{-2}\left(-\cos
y_0\right)}{\partial_\nu\,P_\nu^{-1}\,\left(-\cos y_0\right)}
\,Q_\nu\left(\cosh\chi\right),\label{series}\\
&&P_\nu^{-1}\,\left(-\cos y_0\right)=0. \label{condi}
\end{eqnarray}
First of all, we separate the ultralight mode contribution which
reduces to (\ref{zeroas}) in the limit $\ell \rightarrow \infty$.
That leaves the heavier modes.

Now, using the relation (\cite{bateman}, Eq. 3.4 (14))
\begin{equation}\label{chsgn}
P^\mu_\nu\left(-x\right)= P^\mu_\nu\left(x\right)\,
\cos\left(\pi\left(\nu+\mu\right)\right)- \left(2/\pi\right)
\,Q^\mu_\nu\left(x\right)\,
\sin\left(\pi\left(\nu+\mu\right)\right)
\end{equation}
we rewrite the condition (\ref{condi}) as
\begin{equation}\label{newcond}
\frac{\cos\,\pi\nu}{\sin\,\pi\nu}=\frac{2}{\pi}\,
\frac{Q^{-1}_\nu\left(\cos y_0\right)}{P^{-1}_\nu\left(\cos
y_0\right)}\,\,.
\end{equation}
From this and Eq. (\ref{chsgn}) we obtain
\begin{equation}
P_\nu^{-2}\left(-\cos y_0\right) = \frac{2}{\pi}\,
\frac{\sin\,\pi\nu}{P^{-1}_\nu\left(\cos y_0\right)}\,
\left[P_\nu^{-2}\left(\cos y_0\right)\, Q_\nu^{-1}\left(\cos
y_0\right)- P_\nu^{-1}\left(\cos y_0\right)\, Q_\nu^{-2}\left(\cos
y_0\right)\right]
\end{equation}
In a similar fashion, we rewrite (first using (\ref{chsgn}), and
then using (\ref{newcond}) having taken a derivative with respect
to $\nu$)
\begin{eqnarray}
\partial_\nu\,P_\nu^{-1}\left(-\cos
y_0\right)&=&
\frac{2}{\pi}\,\frac{\sin\,\pi\nu}{P^{-1}_\nu\left(\cos
y_0\right)}\,\left\{P^{-1}_\nu\left(\cos
y_0\right)\,\left[\partial_\nu Q_\nu^{-1}\left(\cos
y_0\right)+\frac{\pi^2}{2}\,P_\nu^{-1}\left(\cos
y_0\right)\right]\right. \nonumber\\&& +\left.
Q^{-1}_\nu\left(\cos y_0\right)\,\left[-\partial_\nu
P_\nu^{-1}\left(\cos y_0\right)+2\,Q_\nu^{-1}\left(\cos
y_0\right)\right]\right\}\,\,.
\end{eqnarray}
Thus we get
\begin{equation}
\frac{P_\nu^{-2}\left(-\cos
y_0\right)}{\partial_\nu\,P_\nu^{-1}\left(-\cos
y_0\right)}=\frac{P_\nu^{-2}\, Q_\nu^{-1}- P_\nu^{-1}\,
Q_\nu^{-2}}{P^{-1}_\nu\,\left[\partial_\nu
Q_\nu^{-1}+\left(\pi^2/2\right)\,P_\nu^{-1}\right]+
Q^{-1}_\nu\,\left[-\partial_\nu P_\nu^{-1}+2\,Q_\nu^{-1}\right]}
\end{equation}
with all the functions at the right hand side evaluated at $+\cos
y_0$.

Now we can go to the limit $\ell\rightarrow \infty$. In this
limit, the masses are related to the values of $\nu$ by
\begin{equation}
m_n\simeq \nu_n/\ell\simeq n/\ell\,,\,\,\,n=1,\,2,\,...\,\,,
\end{equation}
and $m$ becomes a continuous index, so that we can replace
$\sum_n$ by $\int \ell\,dm$. Since $y_0\simeq L/\ell$, $\nu\simeq
m\ell$, $\chi={\cal R}/\ell$ we can rewrite
\begin{eqnarray}
&&P_\nu^{\mu}\left(\cos
y_0\right)=P^\mu_{m\ell}\left(\cos\frac{mL}{m\ell}\right)\,,\quad
Q_\nu^{\mu}\left(\cos
y_0\right)=Q^\mu_{m\ell}\left(\cos\frac{mL}{m\ell}\right)\,,\nonumber\\
&&Q_\nu\left(\cosh
\chi\right)=Q_{m\ell}\left(\cosh\frac{m{\cal R}}{m\ell}\right)\,.
\end{eqnarray}
Then we send $\ell$ to infinity while keeping $m,\,L,\,{\cal R}$
finite, and use the relations (\cite{abram}, Eqs. 9.1.71 \& 72;
9.6.49 \& 50), valid in the limit $\sigma\rightarrow \infty$
\begin{eqnarray}\label{mir}
\sigma^\mu\,P^{-\mu}_\sigma\left(\cos\frac{x}{\sigma}\right)\rightarrow
J_\mu\left(x\right)\,\qquad
\sigma^\mu\,Q^{-\mu}_\sigma\left(\cos\frac{x}{\sigma}\right)\rightarrow
-\frac{\pi}{2}\,Y_\mu\left(x\right)\,\nonumber\\
\sigma^\mu\,P^{-\mu}_\sigma\left(\cosh\frac{x}{\sigma}\right)\rightarrow
I_\mu\left(x\right)\,\qquad
\sigma^\mu\,Q^{-\mu}_\sigma\left(\cosh\frac{x}{\sigma}\right)\rightarrow
{\mathrm e}^{i\mu\pi}\,K_\mu\left(x\right)
\end{eqnarray}
Our sum over massive modes then converges to the integral
\begin{equation}
\frac{p}{2\pi M_5^3}\,\int
\,dm\,\frac{2}{\pi}\,K_0\left(m{\cal R}\right)\,\frac{J_1\left(mL\right)
\,Y_2\left(mL\right)-Y_1\left(mL\right)\,J_2\left(mL\right)}
{J_1\left(mL\right)^2 +Y_1\left(mL\right)^2} \label{eqapp}
\end{equation}
where we have used the fact that the terms of the form
$\partial_\nu P^\mu_\nu$, $\partial_\nu Q^\mu_\nu$, give results
that are proportional to $L/\ell$ and thus vanish in the Minkowski
limit.

Finally, we use the Wronskian relation (\cite{abram}, Eq. 9.1.16)
\begin{equation}
J_{\nu+1}\left(z\right)\,Y_\nu\left(z\right)-Y_{\nu+1}\left(z\right)
\,J_{\nu}\left(z\right)=\frac{2}{\pi\,z}\,\,,
\end{equation}
to simplify (\ref{eqapp}). Including the ultralight mode
contribution in this limit, we finally find
\begin{equation}\label{robe}
f\left(\chi,\,0\right)\big|_{\ell\rightarrow
\infty}=\frac{p}{\pi\,M_5^3\,L}\,\left[\log\frac{{\cal R}}{2L}+1-\frac{1}{2}\,\int
\,\frac{dm}{m}\,\frac{4}{\pi^2}\,
\frac{K_0\left(m{\cal R}\right)}{J_1\left(mL\right)^2+Y_1\left(mL\right)^2}\right]
\end{equation}
which agrees with the result (16) of \cite{roberto}, once we
account for a change in overall normalization.

\end{document}